\documentclass[a4paper,11pt]{article}
\pdfoutput=1
\usepackage{jcappub}
\usepackage{bbold}

\usepackage[english]{babel}
\usepackage[utf8]{inputenc}
\usepackage{amsmath}
\usepackage{color}
\usepackage{amsfonts}
\usepackage{graphicx}
\usepackage{amssymb}
\usepackage{eufrak}
\usepackage{mathtools}

\def\be{\begin{equation}}
\def\ee{\end{equation}}
\def\bea{\begin{eqnarray}}
\def\eea{\end{eqnarray}}
\def\bean{\begin{eqnarray*}}
\def\eean{\end{eqnarray*}}
\def\cd{\cdot}

\def\l {\langle}
\def\re {\rangle}
\def \dd {\partial}

\def \la {\lambda}
\def \La {\Lambda}
\def \De {\Delta}
\def \DH {\Delta_\HI}
\newcommand{\de}{\delta}

\def \al {\alpha}

\def \Om {\Omega}

\def \DW {\Delta w}

\newcommand{\LL}{\mathcal L}

\newcommand{\HH}{\mathcal H}

\newcommand{\Real}{\mathbb{R}}
\newcommand{\bn}{\boldsymbol{n}}
\newcommand{\bv}{\boldsymbol{v}}
\newcommand{\bx}{\boldsymbol{x}}
\newcommand{\bnabla}{\boldsymbol{\nabla}}
\newcommand{\bell}{\boldsymbol{\ell}}
\newcommand{\bal}{\boldsymbol{\alpha}}

\newcommand{\lmax}{\ell_\mathrm{max}}

\newcommand{\HI}{\mathrm{HI}}

\title{Intensity mapping of the 21cm emission: lensing}

\author{Mona Jalilvand,}
\author{Elisabetta Majerotto,}
\author{Ruth Durrer,}
\author{Martin Kunz}

\affiliation{
Universit\'e de Gen\`eve, D\'epartement de Physique Th\'eorique and Centre for Astroparticle Physics,
24 quai Ernest-Ansermet, CH-1211 Gen\`eve 4, Switzerland
}

\emailAdd{mona.jalilvand@unige.ch}
\emailAdd{elisabetta.majerotto@unige.ch}
\emailAdd{ruth.durrer@unige.ch}
\emailAdd{martin.kunz@unige.ch}

\abstract{
In this paper we study lensing of 21cm intensity mapping (IM). Like in the cosmic microwave background (CMB), there is no first order lensing in intensity mapping. The first effects in the power spectrum are therefore of second and third order. Despite this,  lensing of the CMB power spectrum  is an important effect that needs to be taken into account, which motivates the study of the impact of lensing on the IM power spectrum. We  derive a general formula up to third order in perturbation theory including all the terms with two derivatives of the gravitational potential, i.e.\ the dominant terms on sub-Hubble scales. We then show that in intensity mapping there is a new lensing term which is not present in the CMB. We obtain that the signal-to-noise of 21 cm lensing for futuristic surveys like SKA2 is about 10. We find that surveys probing only large scales, $\lmax \lesssim 700$, can safely neglect the lensing of the intensity mapping power spectrum, but that otherwise this effect should be included.
}
\begin{document}

\maketitle

\section{Introduction}
\label{Sec1}
\setcounter{equation}{0}
After the amazing success of Cosmic Microwave Background (CMB) observations, presently major efforts in  cosmology go into the observation and modelling of the distribution of galaxies. As this dataset is  three dimensional, it is potentially much richer and may allow us to study the evolution of the formation of cosmic structure. At high redshifts, the signal from galaxies is however very weak and it is difficult to resolve individual images. Intensity mapping (IM) of a well defined spectral line is a possibility to circumvent this problem.  Intensity mapping (see e.g.\ \cite{Kovetz:2017agg} for a recent overview) is a new technique complementary to galaxy number counts and shear measurements. It will not allow for very high spatial resolution but is mainly sensitive to large scale structure which may reveal the clustering properties of dark energy and other deviations from standard $\La$CDM. 

The most abundant element in the Universe is hydrogen and the 21 centimetre line of the hyperfine structure of neutral hydrogen is well suited for intensity mapping. In the post-reionisation Universe neutral hydrogen (HI) is most abundant in galaxies and protogalaxies and is therefore assumed to be a good tracer of the matter density, see~\cite{Furlanetto:2006jb,Pritchard:2011xb}. Recently, also intensity mapping of the H-alpha line~\cite{2017arXiv171109902S} and of other spectral lines~(see e.g.~\cite{Fonseca:2016qqw,Kovetz:2017agg}) has been proposed. During the complicated reionization process at $6 \lesssim z \lesssim 10$ intensity mapping of hydrogen lines may be used 
to study reionization, but its intensity is not expected to closely follow the matter distribution.
At even higher redshifts, $z>15$ it may again be used to study matter fluctuations, but these low frequencies are beyond the scope of presently planned instruments, except  for HERA~\cite{HERA}.

It has been shown in the past that for galaxy number counts, when going to redshifts of order unity and beyond, lensing by foreground sources cannot be neglected~\cite{Camera:2014sba,Montanari:2015rga,Cardona:2016qxn,Ghosh:2018nsm}. Also in the CMB lensing is very important,  see \cite{Lewis:2006fu,RuthBook,Okamoto:2003zw} and references therein. Observationally it was first detected with the help of cross-correlations in \cite{Smith:2007rg,Hirata:2008cb}, and directly by \cite{Das:2011ak}. In the recent {\em Planck} satellite data the lensing signal is present at very high significance, over $40\sigma$, which permitted the construction of a low resolution map of the lensing potential to the last scattering surface \cite{Ade:2015zua}. It also affects the CMB power spectrum, with a significance of over $10\sigma$ \cite{Ade:2015xua}, and neglecting lensing would lead to a strong bias in parameter inference from CMB data.

It is therefore reasonable to expect that in the future we shall be able to detect lensing in intensity mapping, a prospect that has led to a significant number of papers that study the lensing of HI intensity mapping \cite{Pen:2003yv,Cooray:2003ar,Zahn:2005ap,Mandel:2005xh,Metcalf:2008gq,Zhang:2006fc,Lewis:2007kz,Lu:2007pk,Tashiro:2009ua,Lu:2009je,Kovetz:2012pt,Kovetz:2012jq,Pourtsidou:2013hea,Pourtsidou:2014pra,Pourtsidou:2015mia,Brown:2015ucq,Pritchard:2015fia,Pourtsidou:2015qaa,Pourtsidou:2015ksn,Romeo:2017zwt,Foreman:2018gnv} and of intensity mapping of other spectral features \cite{Croft:2017tur,Metcalf:2017qty,Schaan:2018yeh}.
The advantage of IM lensing with respect to the CMB lensing is that in principle this provides access to the lensing potential not only out to the last scattering surface at $z_*\sim 1000$, but  to arbitrary redshifts outside the reionization window. Most of the lensing signal actually  comes from the redshift range $z\in [2,6]$, which lensing of intensity mapping can capture. 

In this paper we study the impact of lensing on the intensity mapping power spectrum, concentrating mainly on this redshift range ($z\in [2,6]$). 
 As we will see, we obtain new contributions to lensing of 21 cm lines in this redshift range, which imply that not only the trispectrum but also the bispectrum does not vanish. This will modify the quadratic estimator used in much of the literature cited above.

 Even though  the example we present is for the 21 cm line, our results are general and applicable also to other lines observed in intensity mapping e.g. CO or H-$\al$ or Lyman-$\al$ lines, only the appropriate bias changes. The remainder of this paper is structured as follows.
In Section~\ref{Sec2} we first present the formalism to calculate the lensing signal of intensity maps up to third order in cosmological perturbation theory, which is needed to compute the power spectrum up to second order. We take into account all the terms containing the highest number of derivatives of the fully relativistic expression. Apart from the lensing terms these are density and velocity terms  which are also present in a non-relativistic treatment. In Section~\ref{s:num} we  compute the lensing signal numerically and in Section~\ref{SecFish} we discuss its significance using Fisher matrix estimates of the signal-to-noise (S/N) for a future survey out to $z=6$, like e.g.\ SKA~\cite{Maartens:2015mra,Santos:2015bsa}. In Section~\ref{Sec4} we conclude. Some detailed derivations and intermediate results are deferred to Appendices.

\section{Weak Lensing Corrections to Intensity mapping}
\label{Sec2}
\setcounter{equation}{0}
\subsection{Higher order intensity mapping}

It is well known that intensity mapping, like the CMB, does not acquire any corrections from lensing at first order in perturbation theory. Unlike for galaxy number counts, in intensity mapping the increase in the transversal volume due to convergence is exactly compensated by the corresponding increase of the flux.  This is a simple consequence of photon number conservation.  The contribution from lensing that we compute in this paper appears therefore only at second order. 

In this paper we use the metric given by the line element
\be
\mathrm{d}s^2 = a^2(t)\Big[-(1+2\Psi)\mathrm{d}t^2 +(1-2\Phi)\delta_{ij}\mathrm{d}x^i\mathrm{d}x^j\Big]  \label{eq:metric}
\ee
where $\Psi$ and $\Phi$ are the Bardeen potentials. We only keep scalar perturbations, even though at higher order in perturbation theory scalar, vector and tensor perturbations mix, but the vector and tensor perturbations lead to negligible contributions to lensing~\cite{Yamauchi:2012bc,Adamek:2015mna}.

The fully relativistic expression for the first order fractional perturbations for intensity mapping of neutral hydrogen (HI)
has been derived in~\cite{Hall:2012wd}. It is given by.
\bea
\DH(\bn,z) &=&\de_\HI+\frac{1}{\HH(z)}\dd^2_rV 
-3\HH V +\Psi +\HH^{-1}\dot\Phi \nonumber \\
&& +\left(\frac{\dot\HH}{\HH^2} +2-f_{\rm evo}\right)\left(\dd_rV+\Psi+\int_0^{\chi(z)}d\la(\dot\Phi +\dot\Psi)\right)  \,.
\label{e:generalH1}
\eea
This result can also be obtained from the corresponding expression for the number counts~\cite{Bonvin:2011bg,Challinor:2011bk,DiDio:2013bqa}  by setting $2-5s\equiv 0$. Here $\de_\HI$ denotes the HI intensity fractional fluctuations in comoving gauge, $V$ is the potential of the matter velocity, $\bv =\bnabla V$, in longitudinal (Poisson) gauge. $\HH$ is the comoving Hubble parameter, $\HH=aH$ and an overdot is a derivative with respect to conformal time $t$. The parameter $f_{\rm evo}$ parametrizes the physical change in the number density of sources, $f_{\rm evo}= - d\ln(a^3\bar n_\HI)/d\ln a$.

The HI density is $n_bx_\HI$, where $n_b$ is the baryon density and $x_\HI$ is the HI fraction of baryons. If $x_\HI$ is assumed to be independent of the fluctuation amplitude, the HI density fluctuation is simply given by the baryon density fluctuation,
\be\label{e:HI=b}
\de_\HI = \de_b \,.
\ee
To derive~\eqref{e:generalH1} for the HI intensity fluctuations, we have to use that the HI emission intensity is given by the brightness temperature~\cite{Furlanetto:2006jb,Hall:2012wd}
\be\label{e:Tb}
T_b = \frac{3}{32\pi}\frac{h^3A_{10}}{k_BE_{21}}n_bx_\HI(1+z)\left|\frac{d\la}{dz}\,,\right|
\ee
where $h$ is Planck's constant, $k_B$ is Boltzmann's constant, $A_{10}$ is the spontaneous decay rate of 21 cm excitations, and $\la$ is the affine parameter of the incoming photon.
The study of linear perturbations of $T_b$  is presented in detail in~\cite{Hall:2012wd} and it leads to \eqref{e:generalH1}.

\begin{figure}[h]
\begin{center}
\includegraphics[scale=0.45]{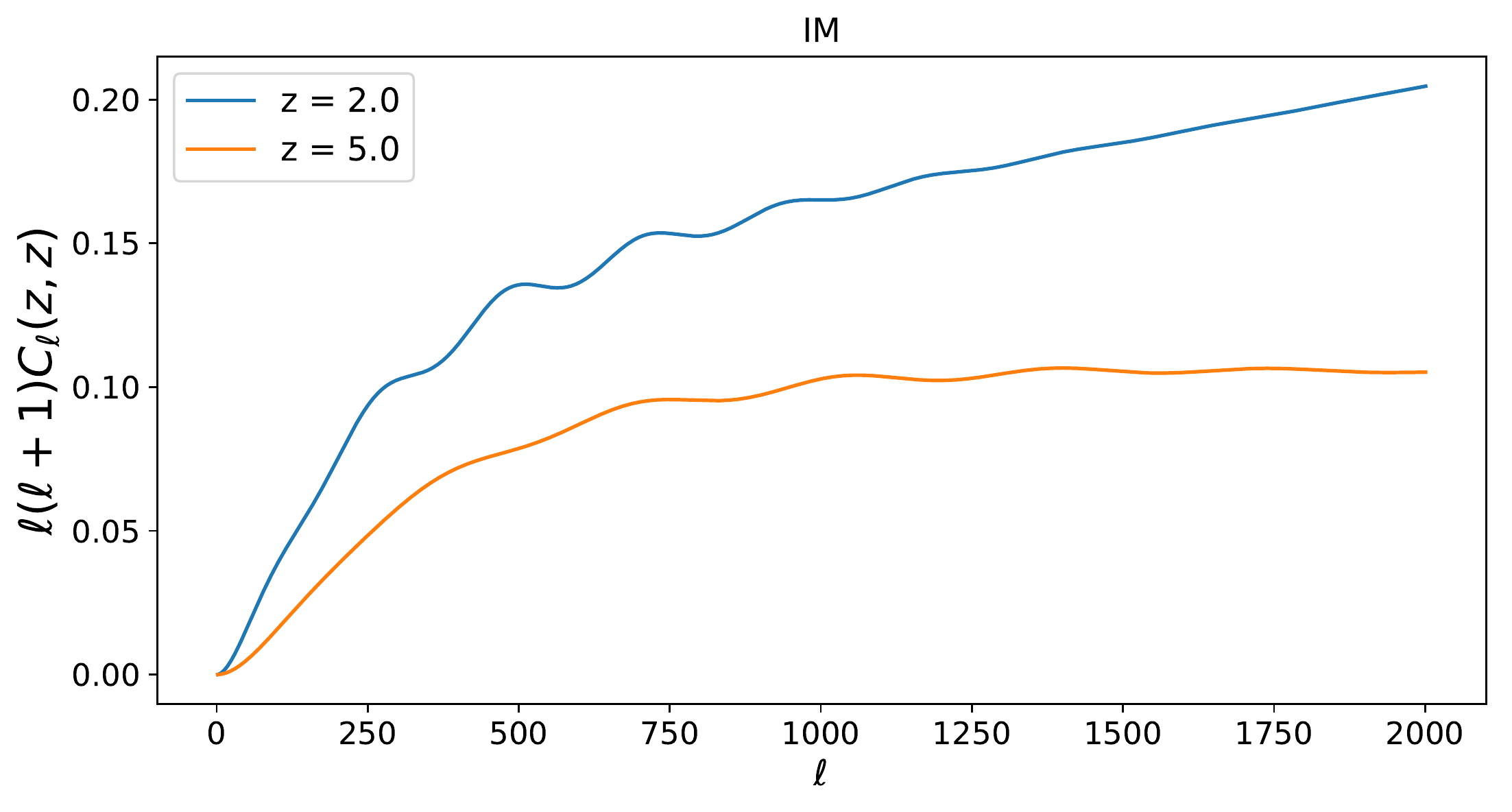}
\includegraphics[scale=0.45]{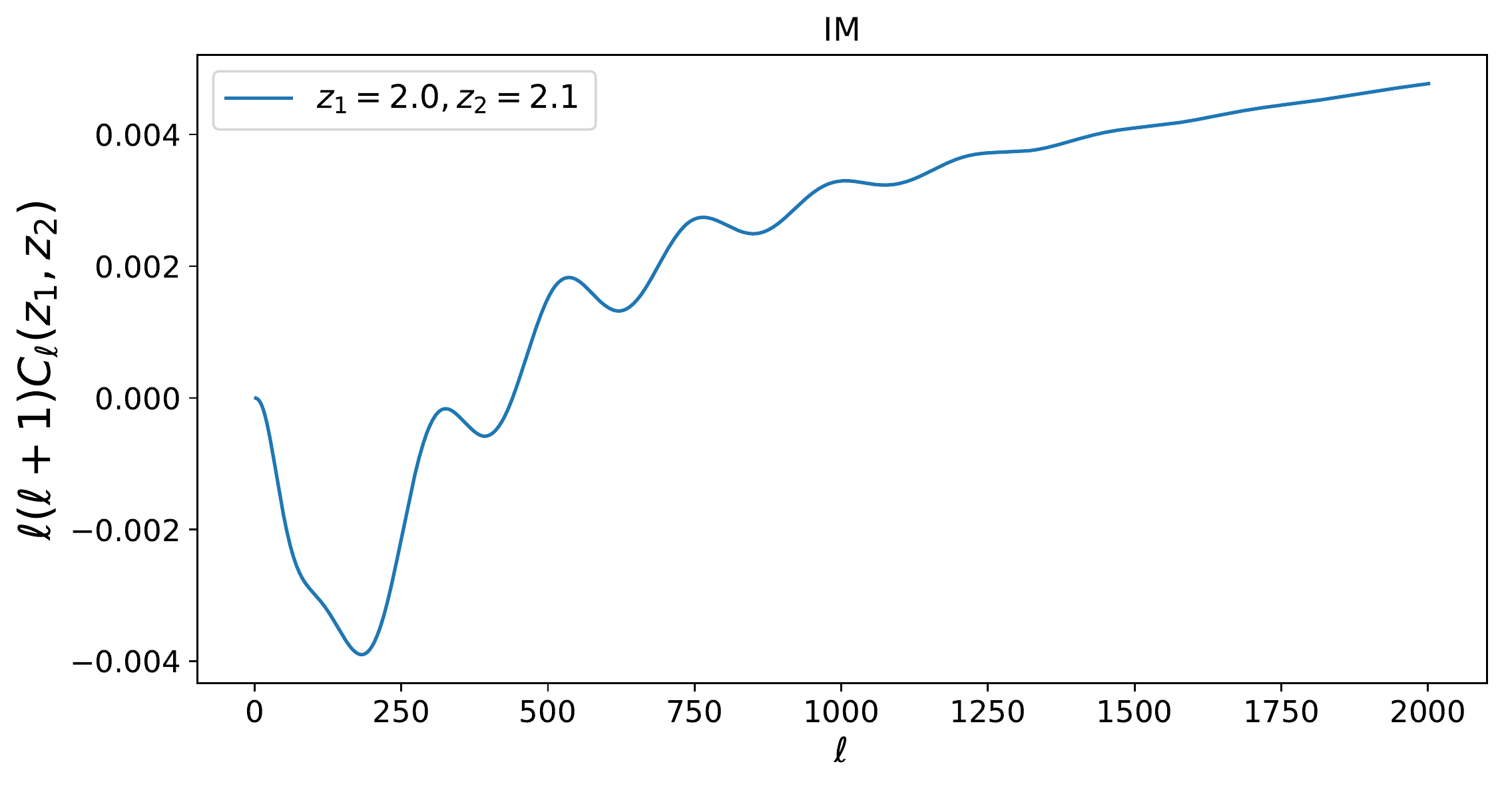}
\includegraphics[scale=0.45]{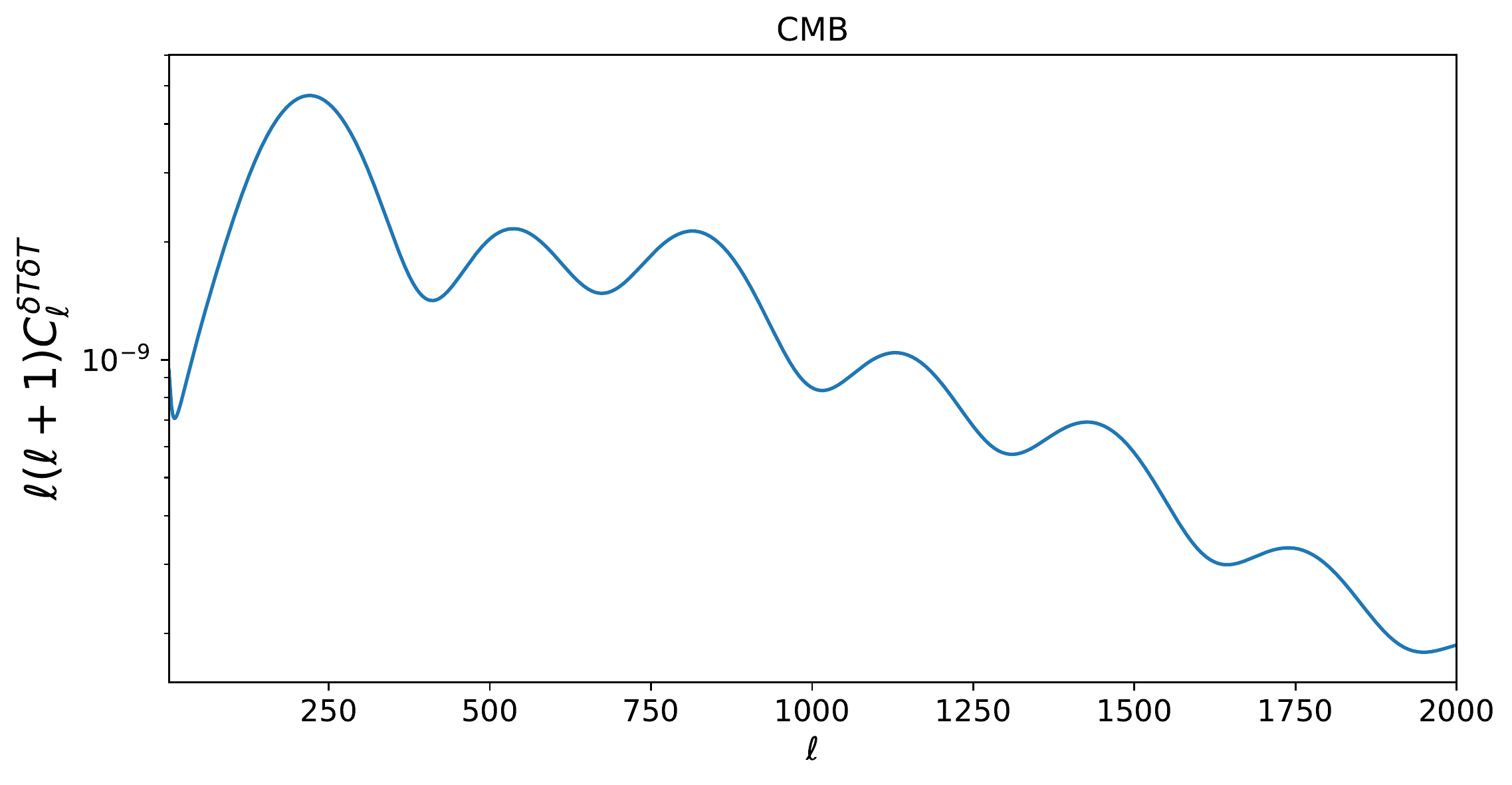}
\end{center}
\caption{Unlensed power spectrum $C_{\ell}^{\Delta \Delta}(z_1,z_2)$. The upper panel shows the auto-power spectrum, with $z_1 = z_2 =2$ and $z_1 = z_2 =5$, computed with a top-hat window function of width $\De z= 0.1$. The middle panel is the cross-power spectrum for $z_1 = 2.0$ and $z_2 = 2.1$, computed with the same window function. Here $\Delta$ includes redshift space distortions and is computed using halofit. The bottom panel shows the unlensed CMB power spectrum.}
\label{fig:powerspectrum}
\end{figure}

In several studies it has been shown that the relativistic large scale effects, i.e.\ all except the first two terms of \eqref{e:generalH1}, are negligible for number counts except at very large scales~\cite{Bonvin:2011bg,DiDio:2013bqa,Umeh:2015gza,Bellini:2017avd}. This is also the case for intensity mapping for the same reason, namely that these effects, for a perturbation with wave number $k$ are suppressed by $(\HH/k)^2$, as we will demonstrate numerically in section~\ref{s:num}. In Fig.~\ref{fig:powerspectrum} we show the unlensed $C_\ell$'s for IM for $z_1=z_2=2$, $z_1=z_2=5$ and  $z_1=2,\;z_2=2.1$ and the unlensed $C_\ell$ for CMB. The baryon acoustic peaks in the IM are clearly visible but they are of course much less pronounced than the corresponding peaks in the CMB. However the peaks and oscillations of the IM cross-power spectrum $(z_1\neq z_2)$ are comparable to those of the CMB. We therefore expect the lensing signal of the IM auto-power spectrum to be much smaller than the one of the CMB while we expect the lensing signal of the cross-spectrum  to be comparable with that of the CMB. Despite this, as we shall see in Section~\ref{s:num}, in  the signal-to-noise calculation the cross-power spectra lensing terms are divided by the much larger auto-power spectra, which significantly reduces their importance. Therefore, we  expect that lensing, which basically leads to a redistribution of power and thereby `smears out' features,  will have less of an effect on IM than it has on the CMB. 

Also fully relativistic second order perturbation expressions have been derived  for both number counts and intensity mapping~\cite{Bertacca:2014dra,Yoo:2014sfa,DiDio:2014lka,Bertacca:2014hwa,DiDio:2015bua,Nielsen:2016ldx}. The full expressions, including all the terms, go over several pages. In Ref.~\cite{Nielsen:2016ldx} the dominant terms, which are relevant on sub-horizon scales, are 
re-derived by a simple, intuitive approach. The second order terms for HI intensity mapping become
\bea\label{e:Hi2}
\DH^{(2)} &=&  \de_\HI^{(2)}  +\HH^{-1} \dd_r^2 V^{(2)}+  \HH^{-1}\delta_\HI^{(1)}\dd_r^2 V^{(1)} + \HH^{-1} \dd_r V^{(1)} \dd_{r} \de_\HI^{(1)}  + \HH^{-2}\dd_r \left( \dd_r V^{(1)} \dd_r^2 V^{(1)}\right)   \nonumber \\
		&&  + \nabla^a\phi^{(1)}\nabla_a \left(\de_\HI^{(1)} +\HH^{-1}\dd_r^2 V^{(1)}\right) 
 \,,	
\eea
where
\be\label{e:lenspot}
\phi^{(n)}(\bn,z) =-2\int_0^{r(z)}dr\frac{r(z)-r}{r(z)r}\Psi_W^{(n)}(r\bn,t(z)) 
\ee
is the $n$th order lensing potential, and $\Psi_W$ is the Weyl potential defined as $\Psi_W = (\Psi + \Phi)/2$, 
see~\cite{RuthBook} for more details on the Weyl potential
and its relation to the lensing potential. 
In \eqref{e:Hi2} the first two terms are simply the second order density and redshift space distortion (RSD).  The expansion contains several more terms that are formally of the same order in $(\HH/k)$. The third term is the mixed product of the two dominant first order terms. The following terms on the first line are the radially displaced first order density and RSD.  For the second of these terms it is important to note that the RSD is given by $\HH^{-1}\dd_r\de z$ with $\de z=\dd_rV$, and it is $\de z$ that we have to shift radially. These Newtonian redshift space distortions have been computed up to third order in Newtonian perturbation theory, see e.g.~\cite{Bernardeau:2001qr}. The terms on the second line are the transversally displaced density and redshift space distortion.
There $\al^a=\nabla^a\phi$ is the deflection  angle. These terms appear in exactly the same way also in the lensing for the CMB temperature~\cite{Lewis:2006fu,RuthBook}.

In this paper we want to determine the effect of lensing on the HI power spectrum. To find the next to leading contribution to the power spectrum, we have to determine also the third order corrections since $\l \DH^{(2)}\DH^{(2)}\re$ is of the same order as $\l \DH^{(1)}\DH^{(3)}\re$. 

To derive the third order correction to $\DH$ we note the following: For an arbitrary perturbation function $F$ on the backward light cone, the second order perturbation contains the dominant terms
\be\label{e:2nd-gen}
[F]^{(2)}(\bn,z) = [F(\bn+\de\bn,z+\de z)]^{(2)}=  \left[F^{(2)}+ \HH^{-1} \dd_r v^{(1)} \dd_{r} F^{(1)}+ \nabla^a\phi^{(1)}\nabla_a F^{(1)}\right](r(z)\bn). 
\ee
Applying this to $F_\HI = \de_\HI +\HH^{-1} \dd_r\de z$ using $F^{(2)}_\HI = \de^{(2)}_\HI +\HH^{-1} \dd_r\de z^{(2)} +\de^{(1)}_\HI\HH^{-1} \dd_r \de z^{(1)} $ with $\de z=\dd_r V$ yields \eqref{e:Hi2}. More details can be found in~\cite{Nielsen:2016ldx}.
At third order the corresponding equation is
\bea
[F(\bn  +\de\bn,z+\de z )]^{(3)} &=&
 F^{(3)}+\HH^{-1} \dd_r V^{(1)} \dd_{r} F^{(2)} + \nabla^a\phi^{(1)}\nabla_a F^{(2)} \nonumber\\
&&+ \HH^{-1} \left[ \dd_r V^{(2)} + \HH^{-1} \dd_r V^{(1)} \dd_r^2 V^{(1)} + \nabla^a\phi^{(1)}\nabla_a \dd_r V^{(1)} \right] \dd_{r} F^{(1)}  \nonumber\\
&&+ \left[ \nabla^a\phi^{(2)} -2 \int_0^{r(z)}dr\frac{r(z)-r}{r(z)r} \nabla^b\phi^{(1)}\nabla_b \nabla^a\Psi_W^{(1)} \right]
\nabla_a F^{(1)}\nonumber \\
&&+ \frac{1}{2} \left[ \HH^{-2} (\dd_r V^{(1)})^2 \dd_{r}^2 F^{(1)}  + \nabla^a\phi^{(1)}\nabla^b\phi^{(1)}\nabla_b\nabla_a F^{(1)}\right] \nonumber \\
&&+ \HH^{-1}\dd_r V^{(1)} \nabla^a\phi^{(1)} \dd_r\nabla_a F^{(1)}  \, . 
\label{e:F3}
\eea
On the right hand side all quantities are now evaluated at the unperturbed positions.

Here we have inserted the higher order `bare' radial and transversal shifts,
\begin{align}
 \de r^{(n)} =H^{-1} \de z^{(n)} &= \HH^{-1}\dd_r V^{(n)} \nonumber \\
\al^{(n)a} &= -2 \int_0^{r(z)}dr\frac{r(z)-r}{r(z)r} \nabla^a \Psi_W^{(n)} \equiv \nabla^a\phi^{(n)} \,.\nonumber
\end{align}
The integral on the third line of \eqref{e:F3} is a so called `post Born term' which takes into account that the second order calculation of the 
deflection angle requires the evaluation of the lensing potential along the perturbed trajectory.

Applying this to intensity mapping yields
\bea
\DH^{(3)} &=& \de^{(3)}_\HI +\HH^{-1}\dd_r^2V^{(3)} +\HH^{-1} \dd_r V^{(1)} \dd_{r}  \de_\HI^{(2)} + \nabla^a\phi^{(1)}\nabla_a \de_\HI^{(2)} \nonumber\\
&&+ \HH^{-1} \left[ \dd_r V^{(2)} + \HH^{-1} \dd_r V^{(1)} \dd_r^2 V^{(1)} + \nabla^a\phi^{(1)}\nabla_a \dd_r V^{(1)} \right] \dd_{r}\de_\HI^{(1)}  \nonumber\\
&&+ \left[ \nabla^a\phi^{(2)} -2 \int_0^{r(z)}dr\frac{r(z)-r}{r(z)r} \nabla^b\phi^{(1)}\nabla_b \nabla^a\Psi_W^{(1)} \right]
\nabla_a \de_\HI^{(1)}\nonumber \\
&&+ \frac{1}{2} \left[ \HH^{-2} (\dd_r V^{(1)})^2 \dd_{r}^2\de_\HI^{(1)}  + \nabla^a\phi^{(1)}\nabla^b\phi^{(1)}\nabla_b\nabla_a \de_\HI^{(1)}\right] \nonumber \\
&&+ \HH^{-1}\dd_r V^{(1)} \nabla^a\phi^{(1)} \dd_r\nabla_a\de_\HI^{(1)}
+\HH^{-1} \dd_r \left\{\HH^{-1} \dd_r V^{(1)} \dd^2_{r} V^{(2)} + \nabla^a\phi^{(1)}\nabla_a \dd_rV^{(2)}\right. \nonumber\\
&&+ \HH^{-1} \left[ \dd_r V^{(2)} + \HH^{-1} \dd_r V^{(1)} \dd_r^2 V^{(1)} + \nabla^a\phi^{(1)}\nabla_a \dd_r V^{(1)} \right]\dd^2_rV^{(1)}\nonumber\\
&&+ \left[ \nabla^a\phi^{(2)} -2 \int_0^{r(z)}dr\frac{r(z)-r}{r(z)r} \nabla^b\phi^{(1)}\nabla_b \nabla^a\Psi_W^{(1)} \right]
\nabla_a  \dd_rV^{(1)}\nonumber \\
&&+ \frac{1}{2} \left[ \HH^{-2} (\dd_r V^{(1)})^2 \dd_{r}^3V^{(1)}  + \nabla^a\phi^{(1)}\nabla^b\phi^{(1)}\nabla_b\nabla_a \dd_rV^{(1)}\right] \nonumber \\
&&\left.+ \HH^{-1}\dd_r V^{(1)} \nabla^a\phi^{(1)} \nabla_a\dd^2_rV^{(1)}\right\} \,.
\label{e:DHi3}
\eea 
Here we neglect gravitational potential and Doppler terms since they are suppressed by roughly a factor $(\HH/k)^2$ with respect to the Newtonian terms and the lensing contributions which we consider. The Newtonian terms, i.e.\ terms including only $\de$ and $V$ in a Fourier-redshift space analysis, which relies on small angles with respect to a common like of sight, are found in~\cite{Bernardeau:2001qr}.  We have checked that when reducing our expressions to this situation they agree with~\cite{Bernardeau:2001qr} (Eqs. (610) to (613)). Formulas up to second order that consider these terms are found in Refs.~\cite{Bertacca:2014wga,DiDio:2014lka,Bertacca:2014hwa}. In Section~\ref{s:large-scales} we show that already the first order contribution of the relativistic terms at $\ell\gtrsim 100$ is significantly smaller than the lensing contribution discussed here.
 
Even though these are only the Newtonian gravity and lensing terms, their number is  already quite significant. In the analysis of the CMB temperature there are two major simplifications which occur. Firstly, the velocity and density fluctuations at the surface of last scattering are quite small so that
higher order contributions to them may be neglected. Furthermore,  lensing mainly comes from redshifts much smaller than the redshift of last scattering, $z_* \simeq 1080$, so that it is reasonable to neglect correlations of $\De T/T$ and the lensing potential. For this reason, lines 3 and 7 in \eqref{e:DHi3} do not contribute since they lead to expectation values of vectors which vanish for reasons of statistical isotropy. For intensity mapping at redshifts $0.1<z<6$ these simplifications are not justified and in principle the full expression \eqref{e:DHi3} has to be considered for the power spectrum beyond leading order.

\subsection{The lensing terms}
As announced in the introduction, in this work we want to study the effects of lensing. Hence we ignore the radial displacements. We shall simply take into account the higher order density and RSD corrections by replacing $\de_\HI$ and $V$ by their `halofit'~\cite{Takahashi:2012em} approximations. Furthermore, it is well known that $\Psi_W^{(2)}\ll \Psi_W^{(1)}$. We therefore also neglect terms containing the higher order
`bare' lensing potential, but we do consider post-Born terms.

Taking into account  only lensing re-mapping we then find up to third order the following expression for $\DH$:
\bea
\DH^{\rm lens} &=&  \DH   + \nabla^a\phi\nabla_a\DH + \frac{1}{2}\nabla^a\phi\nabla^b\phi\nabla_a\nabla_b\DH   
-2\left[\!\int_0^{r(z)}\!\!dr\frac{r(z)-r}{r(z)r} \nabla^b\phi\nabla_b \nabla^a\Psi_W\right]
\!\nabla_a\DH\,. \nonumber\\ && 
\label{e:DHlens}
\eea
Here $\phi$ and $\Psi_W$ denote the lensing and Weyl potentials at first order respectively while for $\DH\equiv \de_\HI +\HH^{-1}\dd_r^2V$ we consider the halofit approximation.
Apart from the last term, the post-Born term, these are exactly the same expressions as the ones obtained for CMB lensing~\cite{RuthBook}.

The post-Born term does not contribute to the CMB for the following reason: If correlations between the lensing potential and $\DH$ can be neglected, this last term yields a second-order contribution of the form
$$ \left\l B^a\right\re\left\l \DH\nabla_a\DH\right\re$$
in the power spectrum, which vanishes due to statistical isotropy: in a statistically isotropic spacetime every expectation value of a spatial vector must vanish. Here we have used Wick's theorem for $\l \DH^{\rm lens}\DH^{\rm lens}\re$, assuming that primordial perturbations are Gaussian.

For HI intensity mapping, we evaluate $\DH$ and $\phi$ at nearly the same redshifts, hence  correlation of $\DH$ and $\phi$ cannot  be  neglected.
For this reason, contrary to the CMB, lensing also induces a bispectrum in IM, with leading contribution 
\be
\left\l \DH \DH \nabla^a\phi\nabla_a\DH \right\re = \left\l \DH \DH \right\re \left\l \nabla^a\phi\nabla_a\DH \right\re \, .
\ee

The IM bispectrum is studied in detail in~\cite{DiDio2018}.

\subsection{The lensed HI power spectrum} \label{sec:lensedCl}
In this section we compute the lensed HI power spectrum in the flat sky approximation at next to leading order. This means we include the terms $\l \DH^{(2){\rm lens}}\DH^{(2){\rm lens}}\re$ and $\l \DH^{(1){\rm lens}}\DH^{(3){\rm lens}}\re$. An introduction to the flat sky approximation can be found e.g. in~\cite{RuthBook}.
For two arbitrary variables $a(\bx,z),~b(\bx,z)$ in the flat sky, $\bx\in \Real^2$, we denote the (unitary) 2d Fourier transform by $a(\bell,z),~b(\bell,z)$ and the power spectra are defined by
\be
\l a^*(\bell_1,z_1)b(\bell_2,z_2)\re =\de^2(\bell_1-\bell_2)C^{ab}_{\ell_1}(z_1,z_2) \,. \label{eq:powspec}
\ee
The Dirac-$\de$ is a consequence of statistical isotropy.
For simplicity we often denote $C^{aa}_\ell$ simply by $C^{a}_\ell$ and $C^{\DH\DH}_\ell$ is  denoted by  $C_\ell$.
 The unlensed $C_\ell$s are shown in Fig.~\ref{fig:powerspectrum} both for the case $z_1=z_2$ and for the case $z_1 \neq z_2$. These and all other power spectra in this paper were computed using  {\sc class}\footnote{http://class-code.net/}~\cite{Blas:2011rf,DiDio:2013bqa}.  

Neglecting correlations of $\DH$ with $\phi$, we first recover the terms, which are also present in the CMB and lead to~\cite{Lewis:2006fu,RuthBook}
\bea
\tilde C_\ell(z_1,z_2) &=&  C_\ell(z_1,z_2) +\int\frac{d^2\ell'}{(2\pi)^2}[\bell'\cd(\bell-\bell')]^2C^\phi_{|\bell-\bell'|}(z_1,z_2)C_{\ell'}(z_1,z_2) \nonumber\\
&& -\frac{1}{2}C_\ell(z_1,z_2)\int\frac{d^2\ell'}{(2\pi)^2}
(\bell'\cd\bell)^2\left[C_{\ell'}^\phi(z_1,z_1)+C_{\ell'}^\phi(z_2,z_2)\right]\,.
\label{e:CMB2}
\eea
To this expression we have to add two new contributions: one from combining the first three terms of \eqref{e:DHlens}, which is due to the non-vanishing correlations of $\phi$ and $\DH$ as well as the post-Born term correlated with $\DH$. The detailed derivation of these new terms is given in Appendix~\ref{a:derCl-new}. Here we just present the full result:
\be
C^{\rm lens}_\ell(z_1,z_2) =  C_\ell(z_1,z_2) +\de C_\ell(z_1,z_2) \label{eq:split}
\ee
with
\bea
\de C_\ell(z_1,z_2) &=& \int\frac{d^2\ell'}{(2\pi)^2}[\bell'\cd(\bell-\bell')]^2 C^\phi_{|\bell-\bell'|}(z_1,z_2)C_{\ell'}(z_1,z_2) \nonumber\\
&& -\frac{\ell^2}{2}C_\ell(z_1,z_2)
\left[R^\phi(z_1,z_1)+R^\phi(z_2,z_2)\right] \nonumber\\   &&
+ 3\Om_mH_0^2 \Big[ \int_0^{r_1}dr\frac{(r_1-r)}{r_1}r(1+z)C^{\de_m\De}_{\ell}(z,z_2)R^{\phi\De}(z,z_1) 
 \nonumber\\
 && + \int_0^{r_2}dr\frac{(r_2-r)}{r_2}r(1+z)C^{\de_m\De}_{\ell}(z,z_1)R^{\phi\De}(z,z_2) \Big] \nonumber\\
 &&+ \int\frac{d^2\ell'}{(2\pi)^2}[\bell'\cd(\bell-\bell')]^2 C^{\phi\De}_{|\bell-\bell'|}(z_1,z_2)C^{\De\phi}_{\ell'}(z_1,z_2)\, .
\label{e:Cellens}
\eea
where 
\bea
R^{\phi}(z_1,z_2) &=&\frac{1}{4\pi}\int_0^\infty d\ell \ell^3C_\ell^{\phi}(z_1,z_2) \equiv \frac{1}{2}\langle\bal(z_1)\cd\bal(z_2)\rangle \,,\\
R^{\phi\De}(z_1,z_2) &=&\frac{1}{4\pi}\int_0^\infty d\ell \ell^3C_\ell^{\phi\De}(z_1,z_2) \,. 
\label{eq:Rphidelta}
\eea
The first two lines of Eq.~(\ref{e:Cellens}) are identical to CMB lensing if we set $z_1=z_2=z_*$ (note that $r=r(z)$ for $r$, $r_1$ and $r_2$). The third and fourth terms are new and come from the fact that we do not neglect correlations between the lensing potential and the intensity fluctuations in this case.

The term $\propto\Om_m$ is the post-Born contribution. We have set $\Psi_W=\Psi=\Phi$ and used the Poisson equation to convert $\De\Phi$ into $\de_m$. This is correct for $\La$CDM but might have to be modified when considering different dark energy models and especially for modifications of gravity. In the numerical results discussed below we consider only $\La$CDM. For $z_1<z_2$ we expect the third line of of Eq.~(\ref{e:Cellens}) to be small, since in this case $z$ (the integration variable) is never equal to $z_2$ and $C_{\ell}^{\delta_m \Delta}(z,z_2)$ is very small when $z\neq z_2$. In that case the post-Born term is dominated by the fourth line. For $z_2<z_1$ the situation is reversed, and the third line will be dominant.
  
We expect  the last line of Eq.~(\ref{e:Cellens}) to be always negligible. For $z_1<z_2$, the lensing potential at $z_1$ is weakly correlated with $\De_{\HI}$ at $z_2$, so the term $C_\ell^{\phi\De}(z_1,z_2)$ is very small. For $z_1>z_2$ the term  $C_\ell^{\De\phi}(z_1,z_2)$ is small,  and for $z_1\simeq z_2$, the lensing kernel is small.
These expectations will be investigated in the next section, see Fig.~\ref{fig:negligibleterm}. 

\section{Numerical Results}
\label{s:num}

\subsection{Higher order lensing contribution}
\label{sec:lensing_terms}
In this section we discuss the numerical results for  the different contribution to the lensing term in Eq.~(\ref{e:Cellens}) which,  for clarity, we denote by
\bea
\de {C_\ell}_{[1]} &\simeq& \int\frac{d^2\ell'}{(2\pi)^2}[\bell'\cd(\bell-\bell')]^2  C^\phi_{|\bell-\bell'|}(z_1,z_2)C_{\ell'}(z_1,z_2) \nonumber\\
\de {C_\ell}_{[2]} &=& -\frac{\ell^2}{2}C_\ell(z_1,z_2)
\left[R^\phi(z_1,z_1)+R^\phi(z_2,z_2)\right] \nonumber \\
\de {C_\ell}_{[3]} &=& 3\Om_mH_0^2\left[\int_0^{r_1}dr\frac{(1+z)(r_1-r)r}{r_1}C^{\de_m\De}_{\ell}(z,z_2)R^{\phi\De}(z,z_1)\right. \nonumber\\ && \hspace{1.42cm} \left.+\int_0^{r_2}dr\frac{(1+z)(r_2-r)r}{r_2}C^{\de_m\De}_{\ell}(z,z_1)R^{\phi\De}(z,z_2)\right]\nonumber
\\
\de {C_\ell}_{[4]} &=&\int\frac{d^2\ell'}{(2\pi)^2}[\bell'\cd(\bell-\bell')]^2 C^{\phi\De}_{|\bell-\bell'|}(z_1,z_2)C^{\De\phi}_{\ell'}(z_1,z_2)
\,.
\label{eq.lensing-terms}
\eea

\begin{figure}[h]
\begin{center}
\includegraphics[scale=0.6]{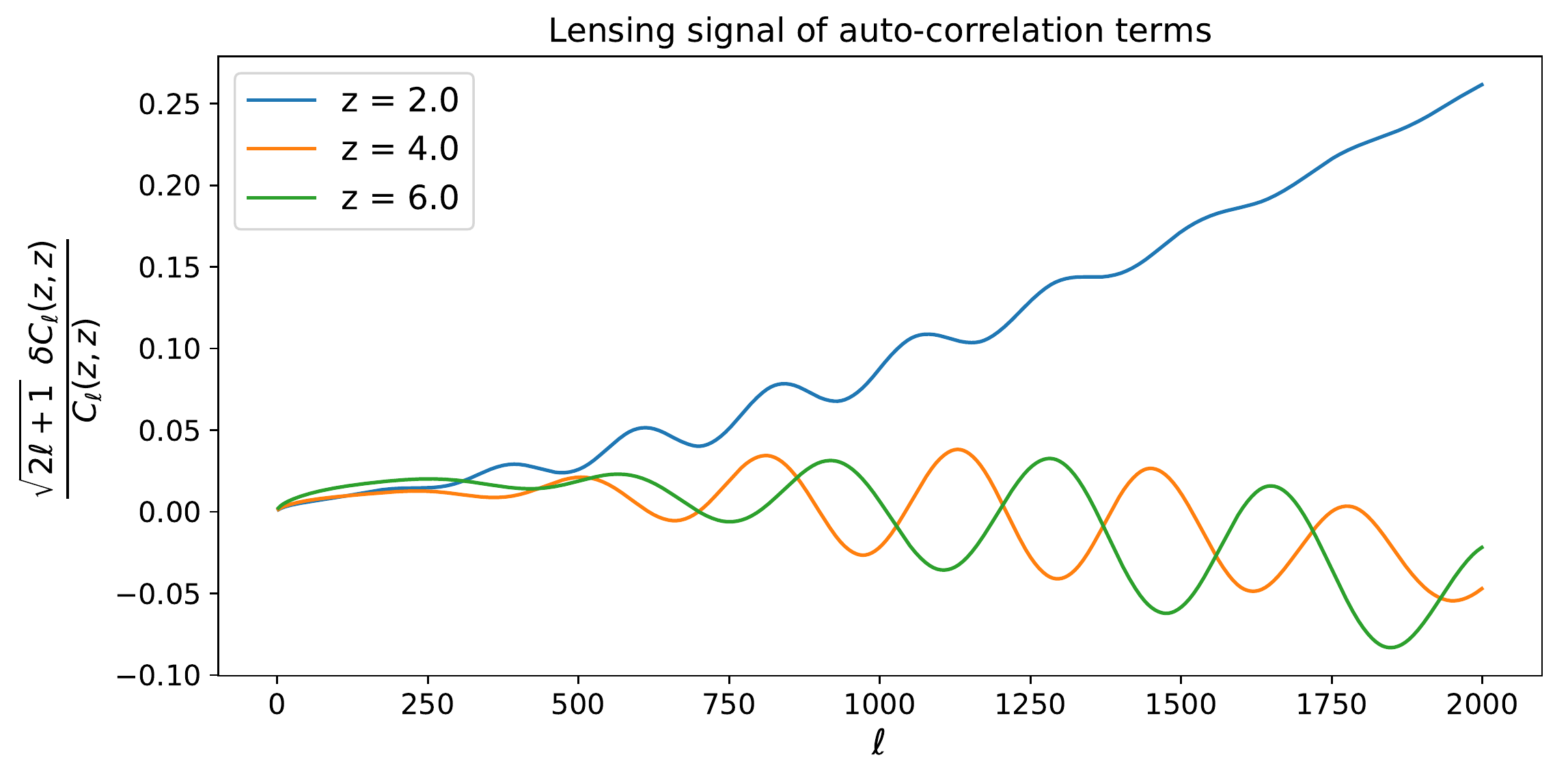}
\includegraphics[scale=0.6]{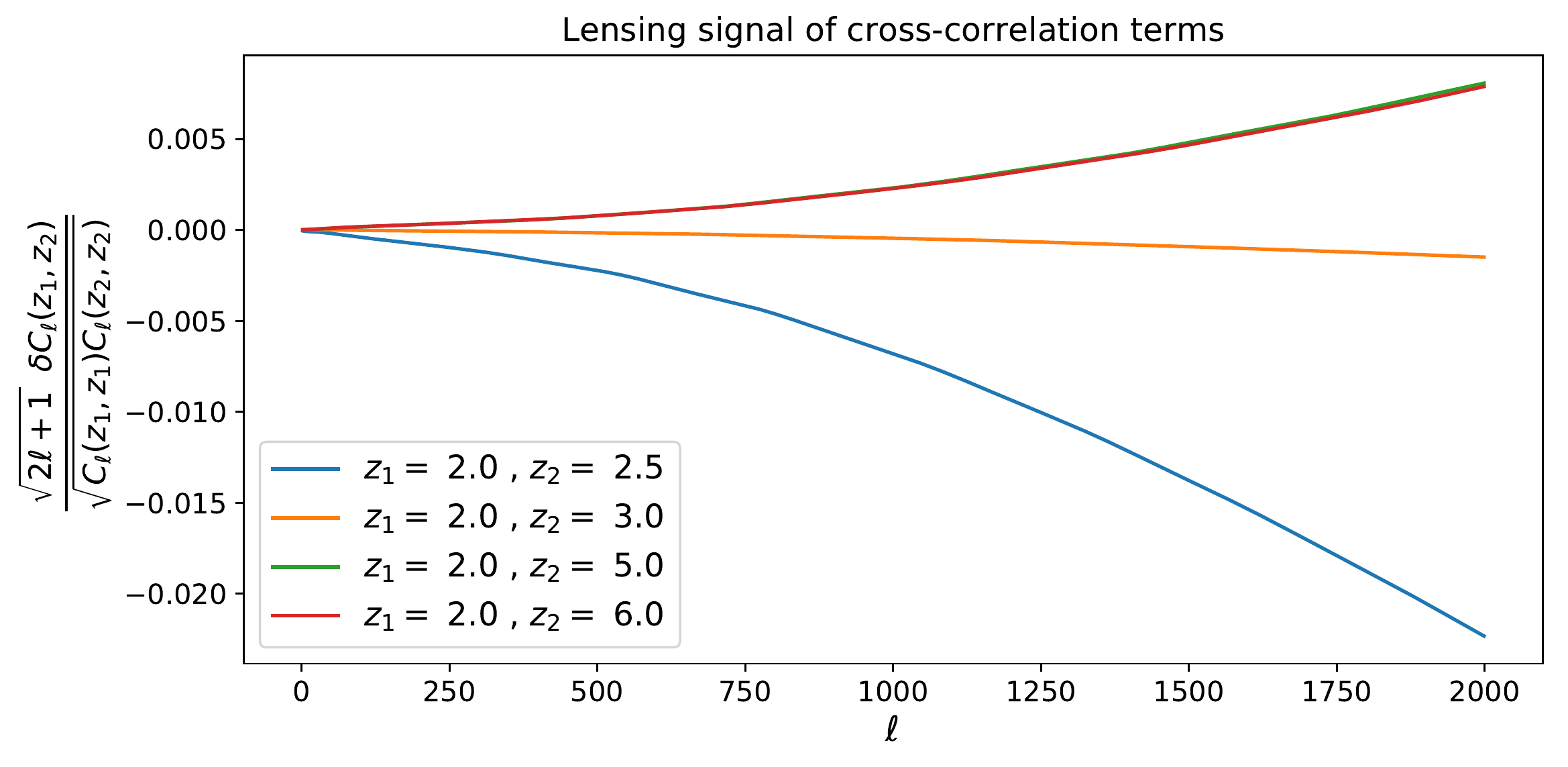}
\end{center}
\caption{ Lensing signal $\de S_\ell(z_1,z_2)$. The upper panel shows the auto-power spectrum lensing terms, $\frac{\sqrt{2\ell+1}\delta C_{\ell} (z,z)} {C_{\ell}(z,z)}$. The bottom panel shows the cross-power spectrum lensing terms, $\frac{\sqrt{2\ell+1}\delta C_{\ell}(z_1,z_2)}{ \sqrt{C_{\ell}(z_1,z_1)C_{\ell}(z_2,z_2)} }$. For auto-power spectra at $z=4$ and $6$, the lensing signal is dominated by $\delta C_{\ell [1]} + \delta C_{\ell[2]}$, which almost cancel each other (see Fig.~\ref{fig:cancellationdevide}), so the resulting signal is oscillating around zero, while the lensing of cross-power spectra and the auto-power spectrum at $z=2$ are dominated by $\delta C_{\ell [3]}$,  which is  smooth. Note also that at fixed $z_1=2$ the lensing cross-power spectrum signal remains constant, i.e. independent of $z_2$ for $z_2\geq 3$.
 }
\label{fig:lensingsignal}
\end{figure}

\begin{figure}[h]
\begin{center}
\includegraphics[scale=0.6]{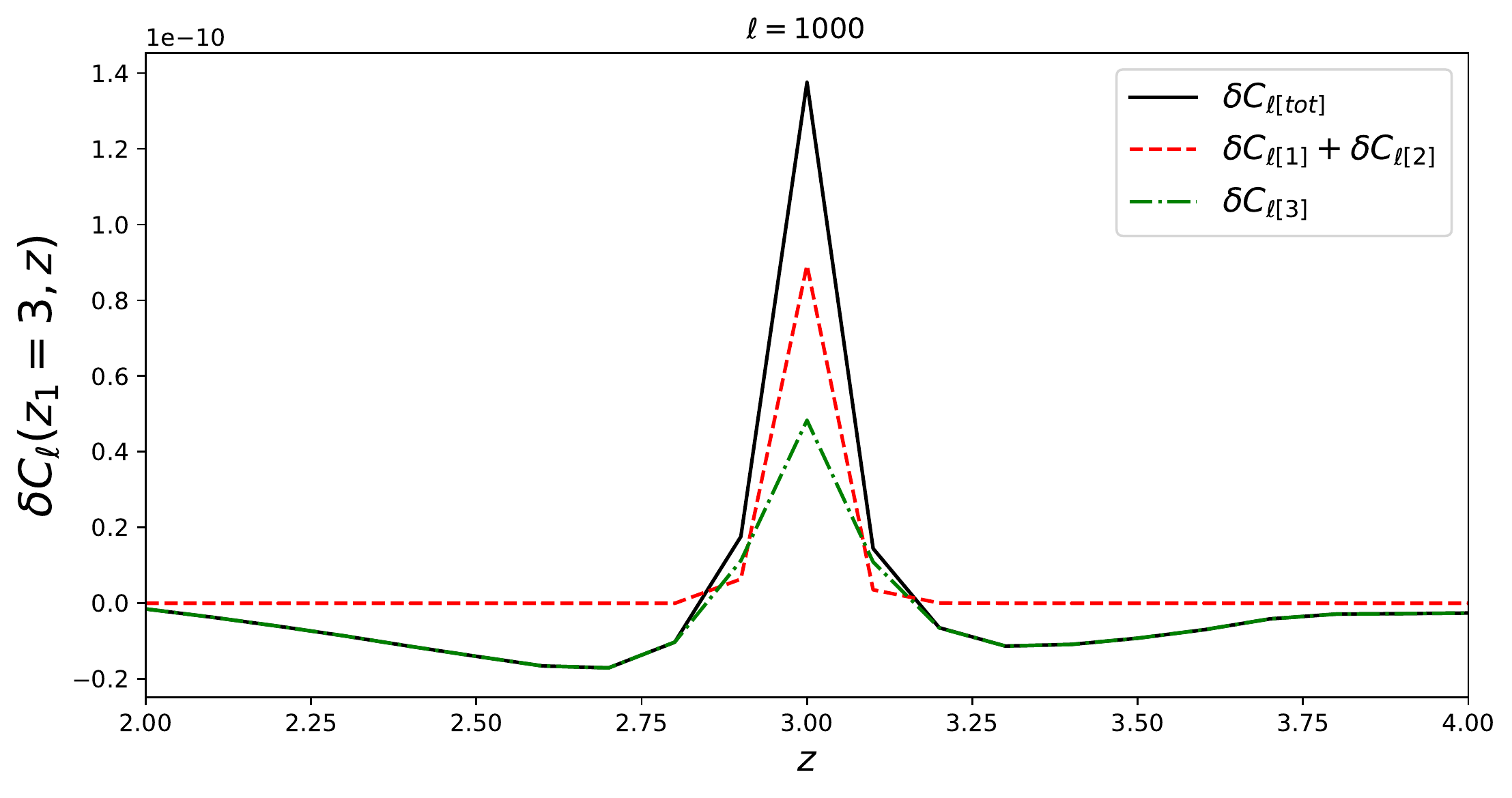}
\end{center}
\caption{Lensing signal  $\de C_\ell^{\rm tot}(z_1,z)$ for $z_1=3$ as a function of $z$, for $\ell = 1000$  (black solid line) and contributions to it from $\delta C_{\ell [1]} + \delta C_{\ell[2]}$ (red dashed line) and $\delta C_{\ell [3]}$ (green dot-dashed line).  While $\delta C_{\ell [1]} + \delta C_{\ell[2]}$ decays very rapidly to zero outside the diagonal, $\delta C_{\ell [3]}$ contributes appreciably to the cross-power spectra until about $\De z\sim 0.75$. }
\label{fig:lensingvsz}
\end{figure}

In Fig.~\ref{fig:lensingsignal} we show $\de S_\ell(z_1,z_2)\equiv\sqrt{2\ell+1}\de C_\ell(z_1,z_2)/\sqrt{C_\ell(z_1)C_\ell(z_2)}$. We prefer this quantity over the relative amplitude $\de C_\ell(z_1,z_2)/C_\ell(z_1,z_2)$ since, as we shall see in the next section, it is this amplitude which enters quadratically in the signal-to-noise calculation. The fact that $\de C_\ell(z_1,z_2)/C_\ell(z_1,z_2)$ can become rather large does not help us to detect the lensing term since the much larger diagonal term $\sqrt{C_\ell(z_1)C_\ell(z_2)}$ also enters in the covariance matrix. We find that the diagonal terms $\de S_\ell(z,z)$ are typically ten times larger than the off-diagonal contributions.  This can also be appreciated in Fig.~\ref{fig:lensingvsz}, which shows the total $\de {C_\ell}(z_1,z)$ as a function of $z$ (black line).  However, the off-diagonal spectra are significantly more numerous if we include many redshift bins.  As we will see in Section~\ref{SecFish} (in particular in Figs.~\ref{fig:SN} and \ref{fig:cumulativeS/N}), for our setup the off-diagonal terms are however still sub-dominant for the total signal-to-noise.  

Interestingly, the total lensing signal decreases with redshift. Even though $\de {C_\ell}_{[1]}+\de {C_\ell}_{[2]}$ is always increasing with redshift, $\de {C_\ell}_{[3]}$ is decreasing. At $z=2$, $\de {C_\ell}_{[3]}$ is larger than $\de {C_\ell}_{[1]}+\de {C_\ell}_{[2]}$, while at $z=3$ the opposite is true on average.

To obtain good accuracy of these small lensing terms we have to determine the corresponding spectra, $C_\ell^{\phi\De}(z_1,z_2)$, $C_\ell^{\phi}(z_1,z_2)$ and $C_\ell^{\De}(z_1,z_2)$ with high accuracy. This requires quite non-standard settings for {\sc class}, which we specify in Appendix~\ref{app:CLASS-settings} for convenience.

\begin{figure}[t]
\begin{center}
\includegraphics[scale=0.6]{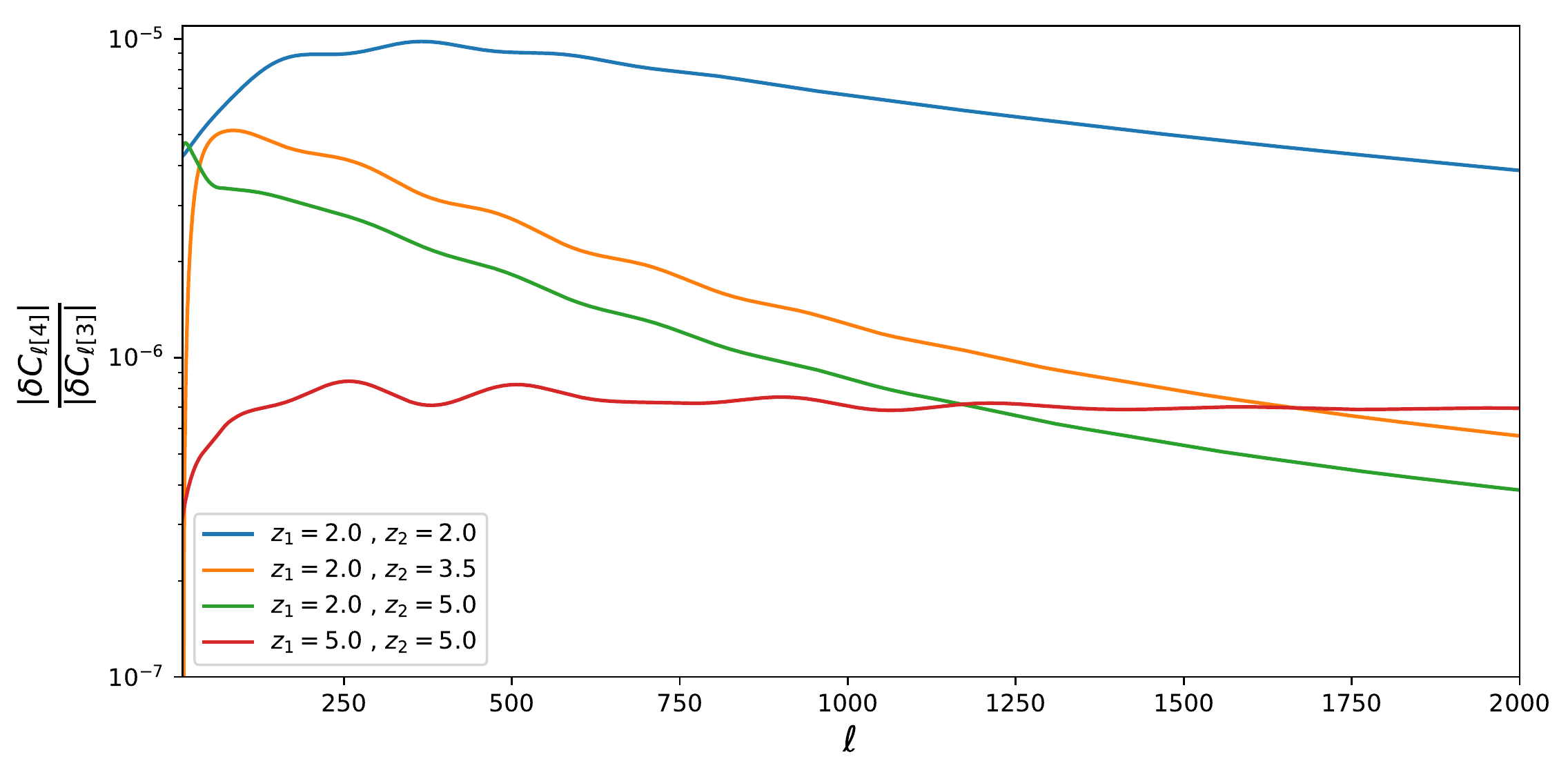}
\end{center}
\caption{Ratio of $|\de {C_\ell}_{[4]}|$ to $ |\delta C_{\ell[3]}|$.  For both auto- and cross-power spectra and for all redshift combinations, this is always less than $10^{-5}$.  } 
\label{fig:negligibleterm}
\end{figure}

As {anticipated} in the previous section, $\de {C_\ell}_{[4]}$ is always highly sub-dominant. In Fig.\ \ref{fig:negligibleterm}, we show the ratio between the absolute value of $\de {C_\ell}_{[4]}$ and   $\de {C_\ell}_{[3]}$ for several redshift pairs. As can be seen $\de {C_\ell}_{[4]}$ is always several orders of magnitude smaller than   the other new term, $\de {C_\ell}_{[3]}$, and therefore also than the total lensing term in Eq.~(\ref{eq.lensing-terms}).\par
\begin{figure}[t]
\begin{center}
\includegraphics[scale=0.6]{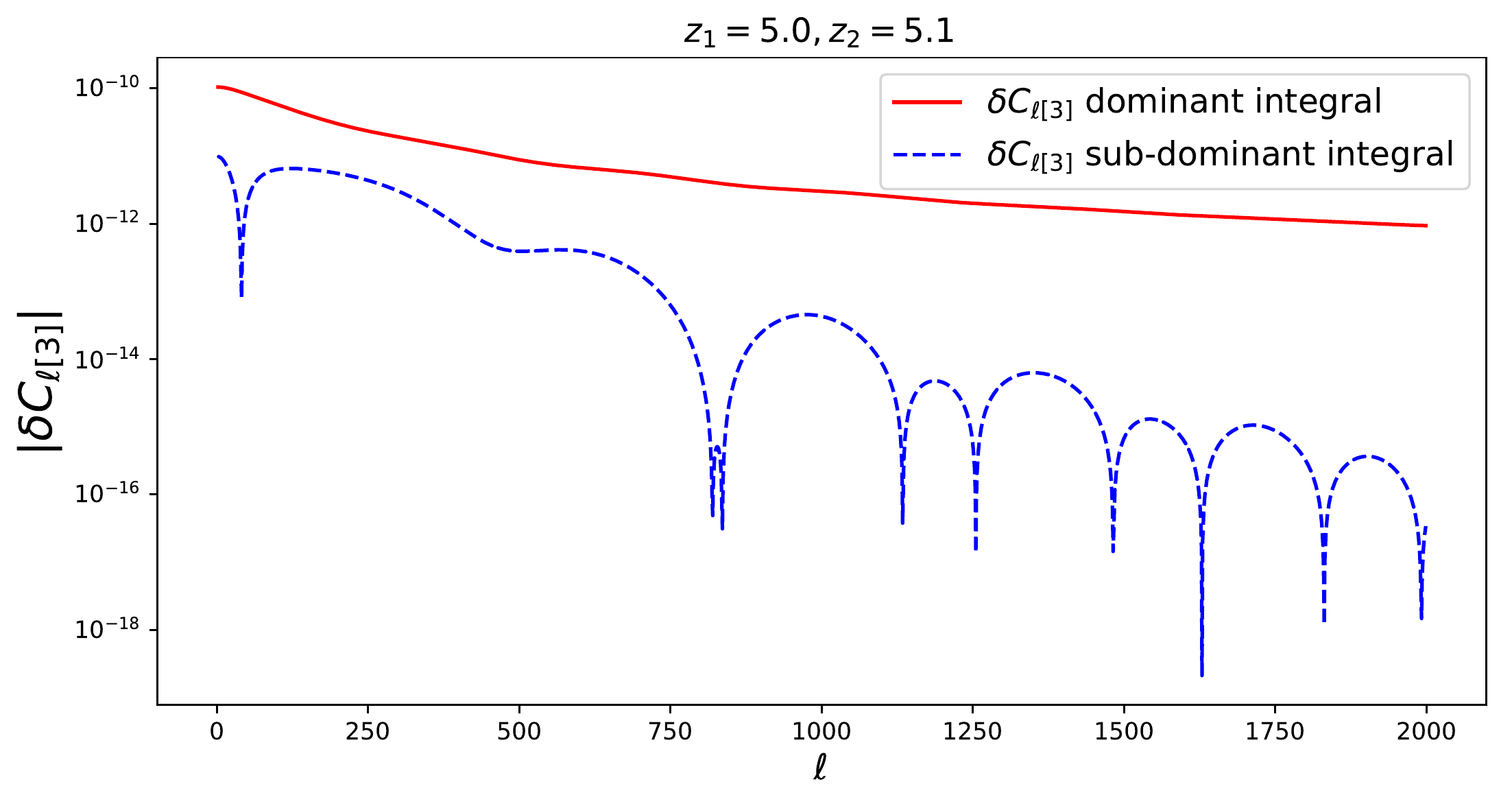}
\includegraphics[scale=0.6]{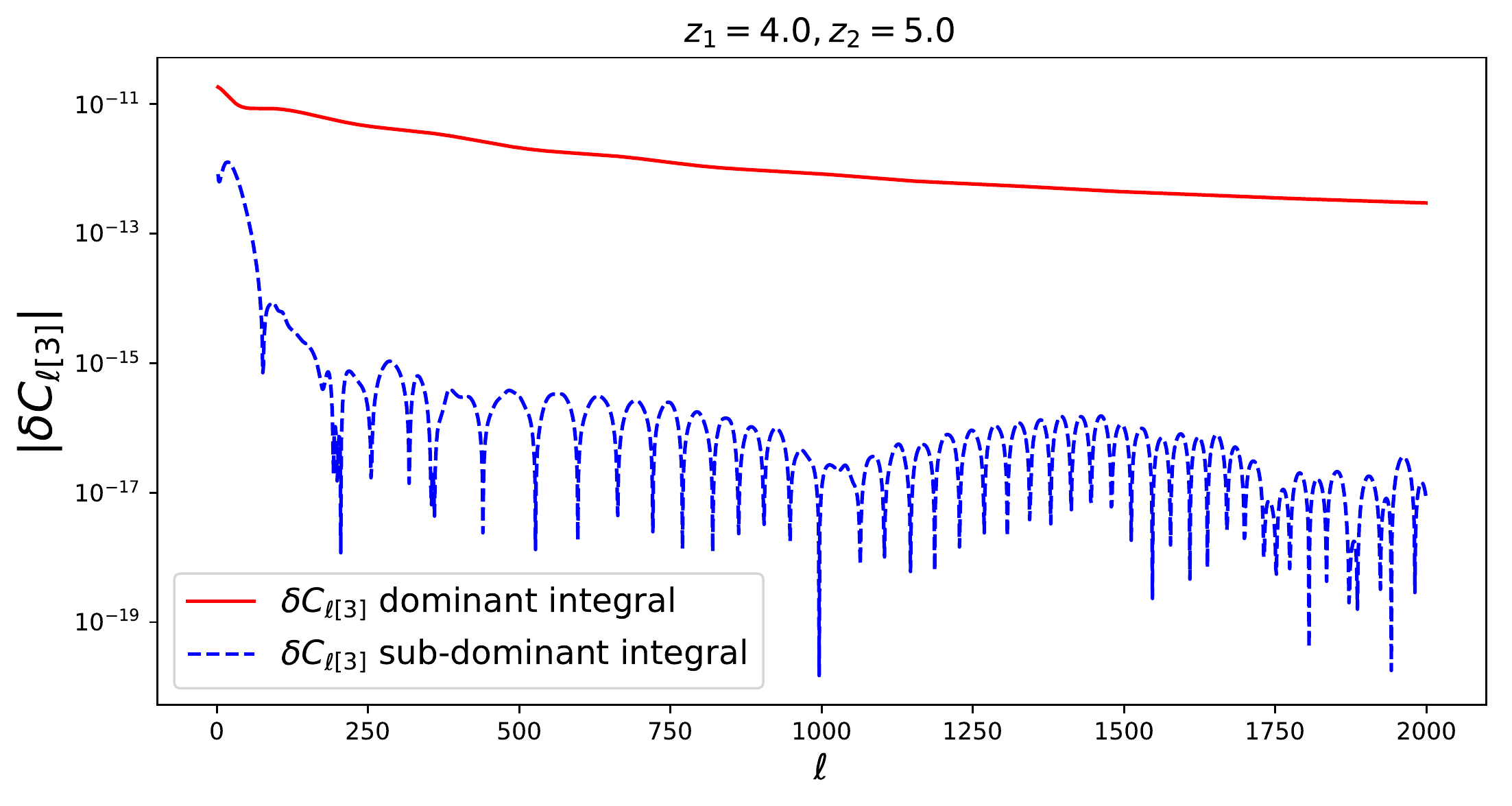}
\end{center}
\caption{ Contribution to $|\de {C_\ell}_{[3]}(z_1,z_2)|$ from the two integrals of Eq.~(\ref{eq.lensing-terms}). The solid red line represents the second integral of which $\de {C_\ell}_{[3]}$ is made of, while the dashed blue line shows its first integral.
The upper panel shows the case of small redshift separations, with $z_1=5$ and $z_2=5.1$, while the lower panel shows the case of large redshift separations, with $z_1=4$ and $z_2=5$.
For the case $z_1 < z_2$ the first integral can be neglected except if $z_1\simeq z_2$. In the case of $z_1\simeq z_2$, the sub-dominant term can contribute up to $\sim 10\%$ for low $- \ell$s. In our calculations we include the sub-dominant term for cross-power spectra which differ by $\Delta z \leq 1$, while we set it to zero when $\Delta z >1$.
}
\label{fig:dominant-term-of-third-term}
\end{figure}

We also note that the term $\de {C_\ell}_{[3]} $ is dominated by the first integral in Eq.~(\ref{eq.lensing-terms}) when $z_1>z_2$ (and by the second integral when $z_2>z_1$). This is shown in Fig.~\ref{fig:dominant-term-of-third-term}. The reason is that the integrand of the second integral contains $C^{\de_m\De}_{\ell}(z,z_1)$, which peaks at $z=z_1$, so that when we integrate only up to $z_2<z_1$, the region where the integrand is largest is not contained in the integration range, and vice versa for $z_2>z_1$.

\begin{figure}[t]
\begin{center}
\includegraphics[scale=0.6]{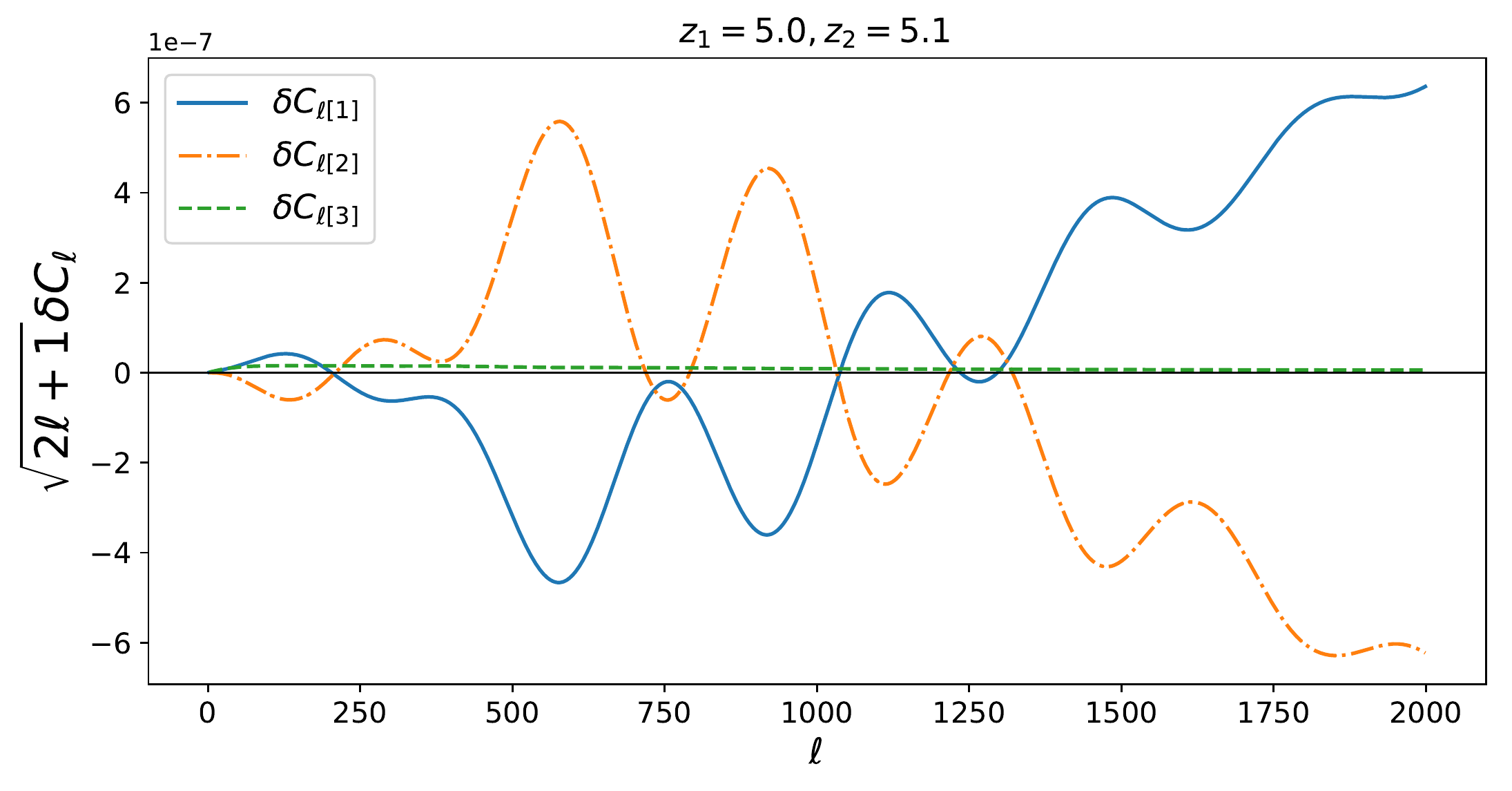}
\includegraphics[scale =0.6]{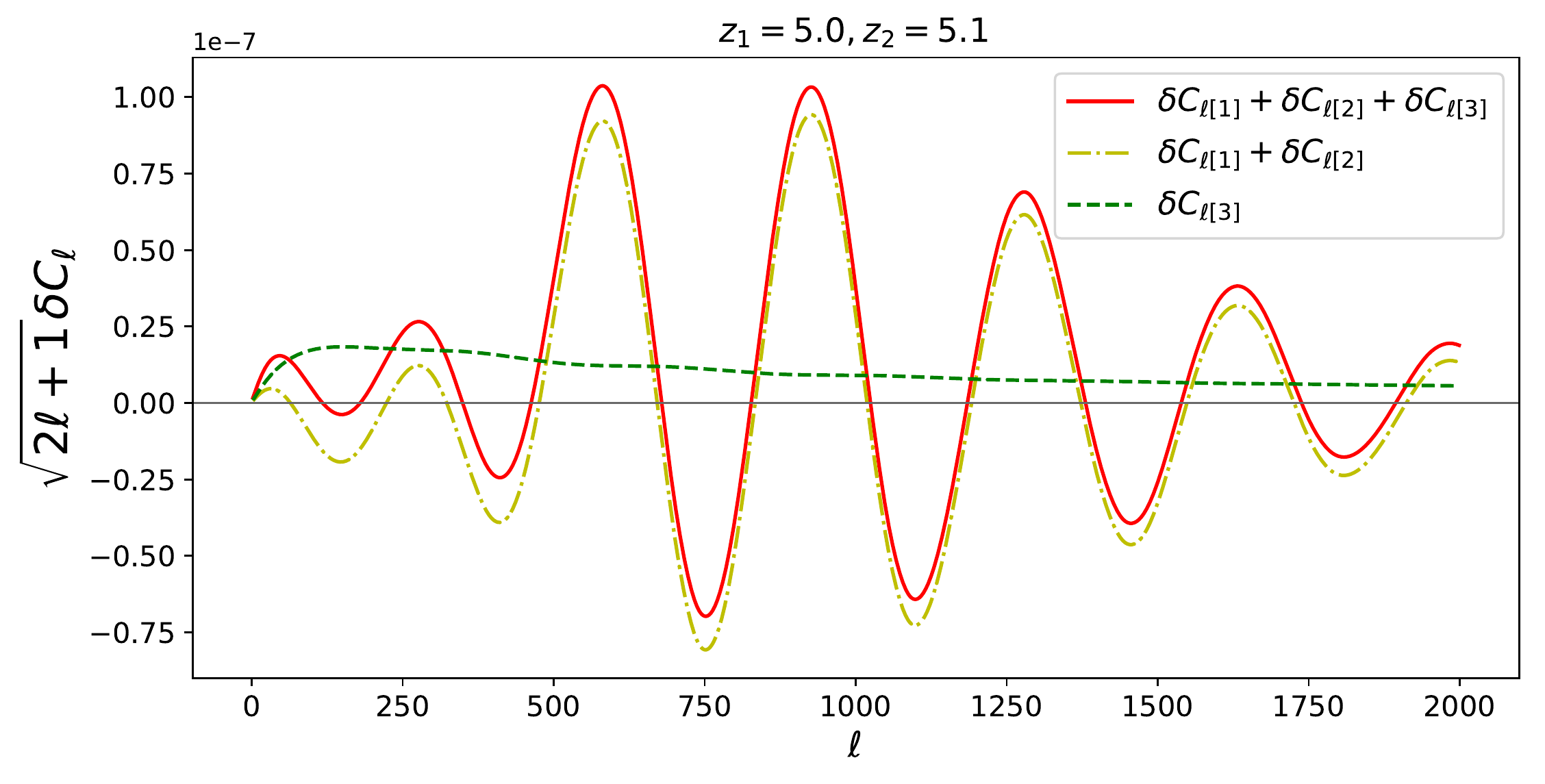}
\end{center}
\caption{
 Contributions of the different lensing terms to the cross-power spectrum $C_\ell(z_1=5.0,z_2=5.1)$.
The top panel shows how $\de {C_\ell}_{[1]}$ (solid blue line)  and $\de {C_\ell}_{[2]}$ (dot-dashed orange line) nearly cancel each other. For this reason in this plot it is not possible to appreciate the contribution from  $\de {C_\ell}_{[3]}$. We therefore show in the bottom panel the total lensing term (solid red line) along with the contribution from $\de {C_\ell}_{[1]}+ \de {C_\ell}_{[2]}$ (dot-dashed yellow line) and $\de {C_\ell}_{[3]}$ (dashed green line). For such small separation $|z_2-z_1| = 0.1$, the dominant contribution comes from $\de {C_\ell}_{[1]}+ \de {C_\ell}_{[2]}$.
} 

\label{fig:lensingterms}
\end{figure}

In Fig.~\ref{fig:lensingterms} we show the contribution of the different lensing terms to the cross-power spectrum $C_\ell(z_1=5.0,z_2=5.1)$.
Comparing the dashed orange and solid blue lines we see that $ \de {C_\ell}_{[1]}$ and $\de {C_\ell}_{[2]}$ cancel almost exactly. This cancellation, which is also present in CMB lensing, is explained in Appendix~\ref{app:cancellation}. 
To obtain $\de {C_\ell}_{[3]}$, shown as a green line in Fig.~\ref{fig:lensingterms}, we cannot simply compute the integral in Eq.~(\ref{eq.lensing-terms})\footnote{We might be tempted to use here the often employed Limber approximation leading to $C_{\ell_2}^{\de_m\DH}(z,z_2) \simeq \de(z-z_2)C_{\ell_2}^{\de_m\DH}(z_2,z_2)$, but  this also leads to $R^{\phi\Delta}(z,z_1) \sim\de(z-z_1)R^{\phi\Delta}(z_1,z_1)$ so that we only have a contribution for $z_1=z_2$. However, we have found that this approximation is not sufficient and we have to include the tail of $R^{\phi\Delta}$. We have checked this numerically for a large number of cases.}, but we have to consider that the observations are averages over finite size redshift bins. This means that instead of observing the power spectrum of $\langle \DH^{*\rm lens}(\bell_1,z_1)\DH^{\rm lens}(\bell_2,z_2) \rangle $ we will observe the power spectrum smoothed over redshift window functions:
\bea \nonumber
&& \langle \DH^{*\rm lens}(\bell_1,z_1) \DH^{\rm lens}(\bell_2,z_2) \rangle^{\rm obs} \\
&&\,= \int dz'W(z',z_2) \int dz'' W(z'',z_1) \langle{\Delta}_{\mathrm{HI}}^{*\rm lens}(\bell_1,z'') {\Delta}_{\mathrm{HI}}^{\rm lens}(\bell_2,z') \rangle\,.
\eea
The details about the calculation of this integral are given in Appendix~\ref{app:third-term-derivation}.

\begin{figure}[t]
\begin{center}
\includegraphics[scale=0.6]{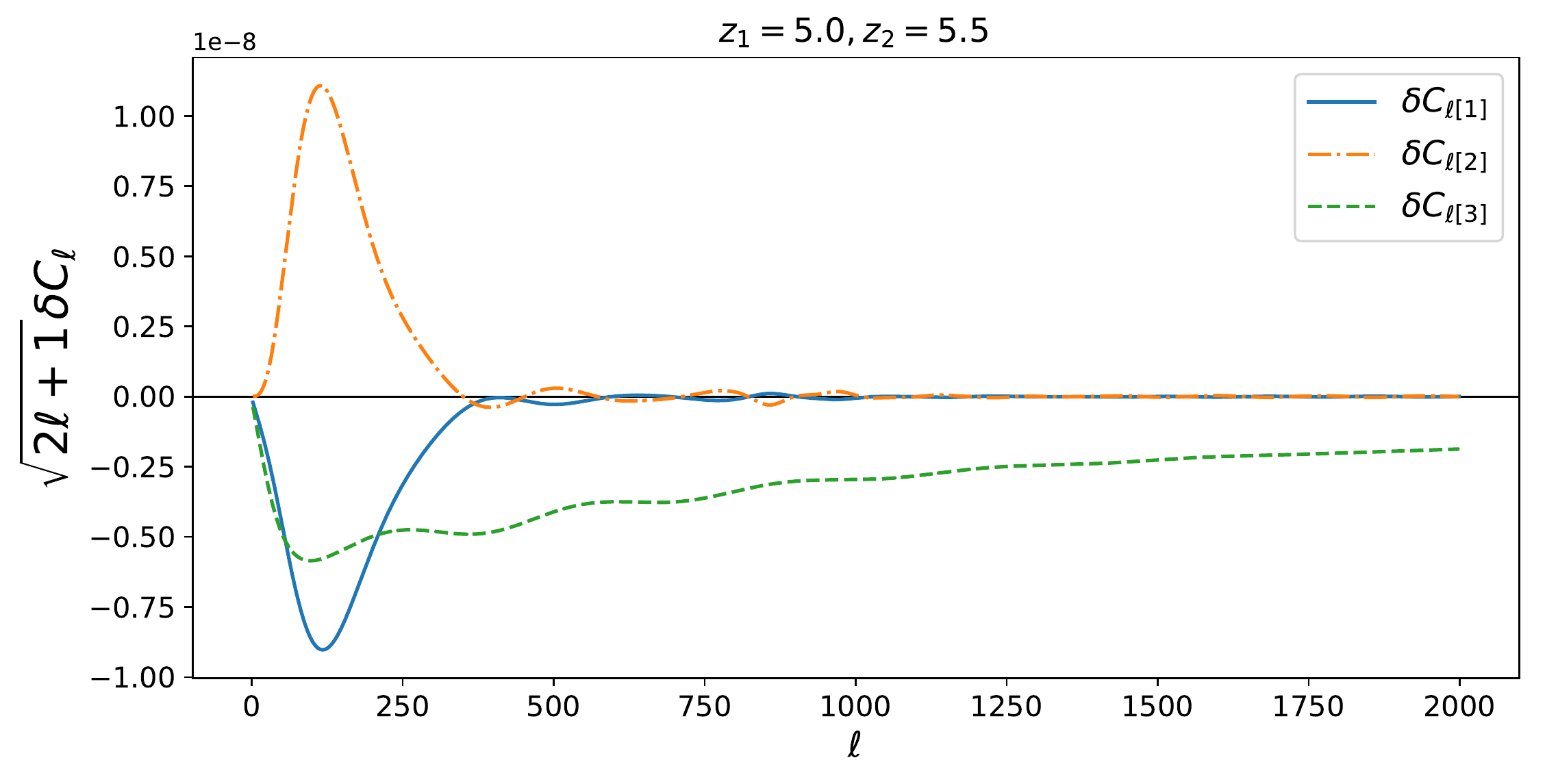}
\includegraphics[scale =0.6]{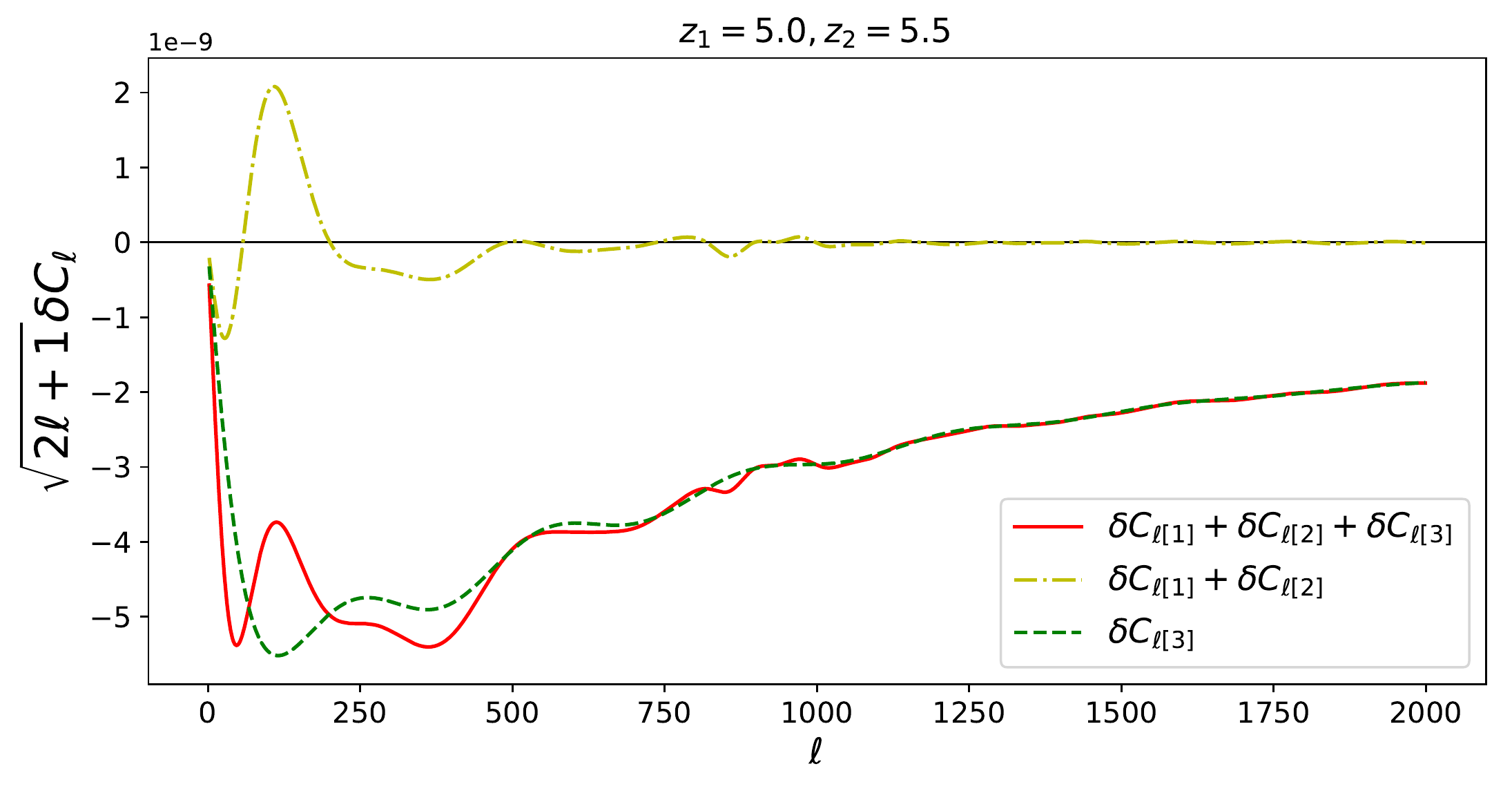}
\end{center}
\caption{Same as Fig.~\ref{fig:lensingterms} but for $z_1=5.0,z_2=5.5$. In this case $\de {C_\ell}_{[3]} $ dominates the total lensing signal. \label{fig:lensingterms2}
}
\end{figure}

 In Fig.~\ref{fig:lensingvsz} we show the total lensing signal $ \de {C_\ell}_{\rm [tot]}(z_1,z)$ (black line) as a function of  $z$ for a fixed $z_1 = 3.0$ and for $\ell = 1000$ , along with the contribution from $ \de {C_\ell}_{[1]}+\de {C_\ell}_{[2]}$ (red dashed line) and from $ \de {C_\ell}_{[3]}$ alone (green dotted line). For equal redshifts, $z=z_1=3$, the contribution of the latter is smaller than the one of the former two terms, but still non-negligible. As the redshift separation increases, the lensing signal becomes smaller and completely dominated by $ \de {C_\ell}_{[3]}$. This is also visible in Fig.\ \ref{fig:lensingterms2}.

\subsection{Comparing lensing terms with gravitational potential terms}\label{s:large-scales}
\begin{figure}
\begin{center}
\includegraphics[scale=0.6]{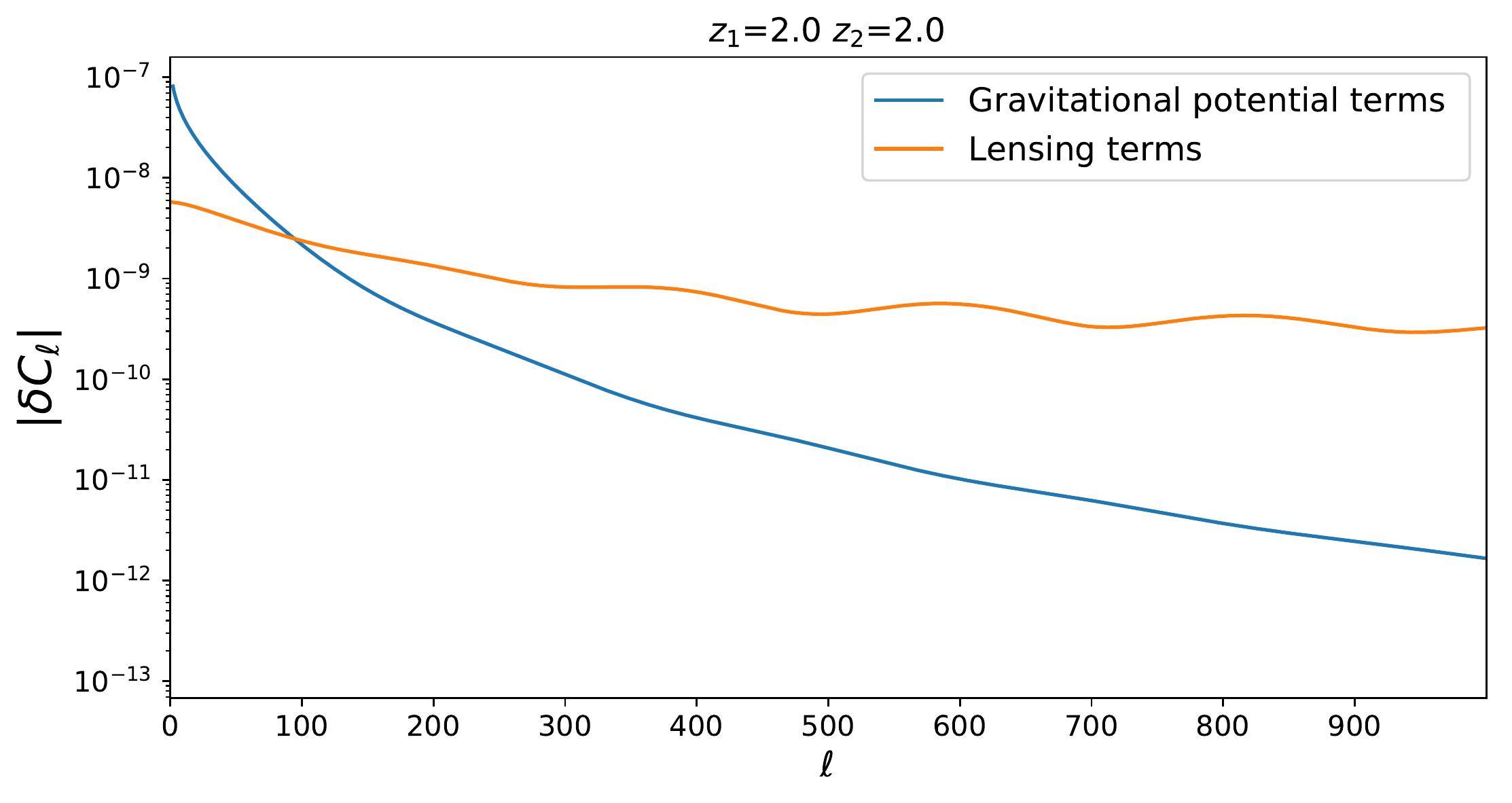}
\includegraphics[scale=0.6]{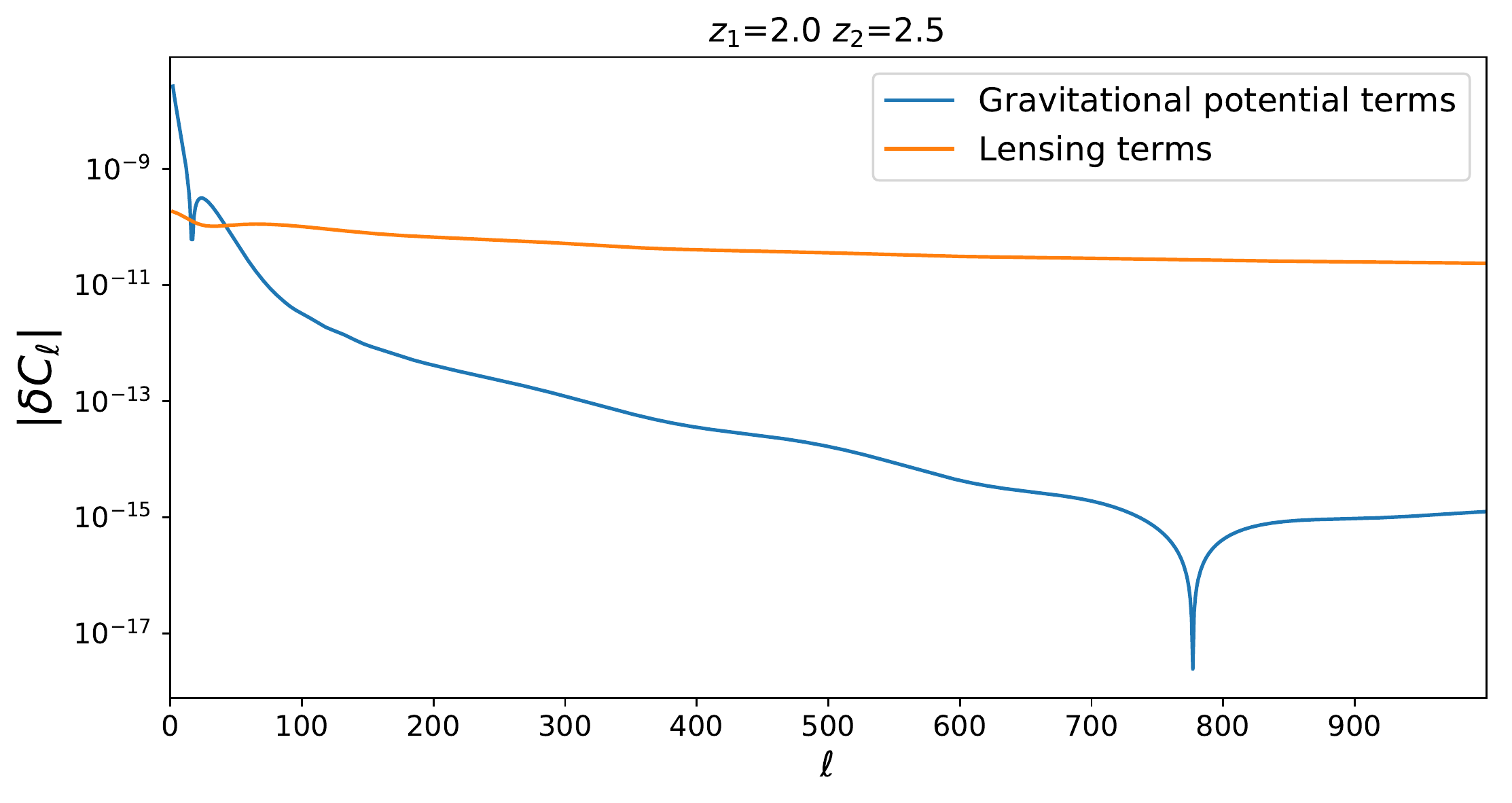}
\end{center}
\caption{Gravitational potential terms (blue lines) and second and third order lensing terms (orange lines) for auto-power spectra, $C_{\ell}(z_1=2,z_2=2)$ (top panel), and for cross-power spectra $C_{\ell}(z_1=2,z_2 = 2.5)$ (bottom panel).
}
\label{fig:grterms}
\end{figure}

In our calculation we have neglected the gravitational potential terms which are suppressed by $(\HH/k)^2$ with respect to density and RSD, but which are present already at first order.
In this section we compare the contribution of second and third order lensing terms with these first order gravitational potential terms, introduced in equation ($2.17$) of ~\cite{DiDio:2013bqa}, containing an integrated Sachs-Wolfe term and terms depending on the gravitational potentials, $\Phi$ and $\Psi$.
Since we found that the lensing contribution to the IM spectra is very small, we want to investigate whether the gravitational potential terms are really smaller than the lensing terms. As lensing builds up with redshift and as relativistic terms decay with $\ell$, we expect the relativistic terms to dominate at low $\ell$ and at low redshift. This is exactly what we find.
 In Fig.~\ref{fig:grterms}, we compare first order gravitational potential terms and second and third order lensing contributions to the power spectrum for two redshift pairs. Lensing dominates over the gravitational potential terms for $\ell\gtrsim 120$,  the exact value depends on redshift.
 
The contribution to cross-correlation spectra depends on the redshift pairs, but the largest value of $\ell$ at which the gravitational potential terms dominate for cross-correlations of $z=2.5$ with other redshifts is at $\ell =150$. As shown in the next section, specifically in Fig.~\ref{fig:SN}, the lowest $\ell$ contribute least to the overall signal-to-noise of the lensing term so that the gravitational potential terms will not significantly affect a detection of IM lensing.

\section{Is  lensing of intensity mapping observable?}
\label{SecFish}

As a first result, the plots in the previous section have shown that lensing is a relatively small effect on the intensity mapping (IM) power spectra, much smaller than in the CMB where it becomes 10\% and more at $\ell>2000$ and even dominates the signal for $\ell>5000$ while for IM power spectra, lensing never contributes more than 0.1\% for $\ell\leq 2000$.  One reason for this is that density fluctuations at lower redshifts, $z<6$ are much larger than CMB anisotropies while the lensing potential is of the same order. But more importantly, on intermediate scales, $100<\ell<4000$ lensing mainly redistributes power and on a relatively smooth signal like the IM power spectrum the effect of a redistribution is  not very significant. This may be different on much smaller scales, $\ell>5000$, but on these scales nonlinearities of the matter distribution are very relevant and our perturbative approach can no longer be trusted. For this reason we restrict ourselves to $\ell<2000$ in this work.

Despite this drawback, the fact that we can observe intensity mapping for a large number of redshifts, as well as the use cross-spectra, neither of which is possible with the CMB, can boost the final signal-to-noise of the effect. Here we study whether a large number of redshift bins and their cross-correlations  is able to compensate for the smallness of the  effect. 

We use a Fisher matrix approach with optimistic assumptions: we assume that shot-noise is not an issue, i.e.\ that our IM spectra are cosmic variance dominated to $\lmax = 2000$, and that we can perform observations from $z_\mathrm{min} = 2$  to $z_\mathrm{max} = 6$. For definiteness we will use a redshift bin width of $\Delta z = 0.1$,  which leaves the bins very weakly correlated and also suppresses the contribution from redshift space distortions, which we consider here only at first order, since we are interested mainly in the lensing effect.

The data vector observed by intensity mapping at a redshift $z$ can be taken to be the coefficients $a_{\ell m}(z)$ of a spherical harmonic transform of the IM map, similar as for the CMB. We assume that at fixed redshift the IM fluctuations form a Gaussian random field on the sphere. Even though the non-linearities and the lensing term render the field slightly non-Gaussian, we neglect this in the calculation of the covariance matrix and the signal-to-noise. The power spectrum is then given, as in Eq.\ (\ref{eq:powspec}), by
\be
\langle a^*_{\ell m}(z_i) a_{\ell' m'}(z_j) \rangle = \delta_{\ell \ell'} \delta_{m m'} C_{\ell}(z_i,z_j)
\ee
where we will write $C_\ell^{ij} = C_{\ell}(z_i,z_j)$. The likelihood for the $a_{\ell m}$ is of the standard multivariate Gaussian form
\be
\ln \LL = -\frac{1}{2} \left[  \sum_{\ell, m} \sum_{ij}  a^{*i}_{\ell m} (C^{-1}_\ell)^{ij} a^j_{\ell m} - \ln \det (C^{ij}_\ell) \right] + \mathrm{const.}
\ee

The theoretical covariance matrix $C^{ij}_\ell$ depends in general on cosmological and other parameters $\theta_\alpha$, and the uncertainty with which we can recover these parameters is given by the Fisher matrix \cite{Tegmark:1996bz}
\be
F_{\alpha\beta} = \left\langle -\frac{\partial^2 \ln \LL}{\partial \theta_\alpha \partial \theta_\beta} \right\rangle \, ,
\ee
specifically, the parameter covariance matrix ${\rm Cov}_{\alpha\beta} \approx F^{-1}_{\alpha\beta}$. The Fisher matrix is given by (see, e.g.~\cite{RuthBook}, Eq. (6.41))
\be
F_{\alpha\beta} = \sum_\ell \frac{2\ell+1}{2} \sum_{a,b,c,d} \left[ (\partial_\alpha C^{ab}_\ell) (C^{-1}_\ell)^{bc} (\partial_\beta C^{cd}_\ell) (C^{-1}_\ell)^{da} \right] \, . \label{eq:fishgen}
\ee
A partial sky coverage can be approximately incorporated by scaling $F_{\alpha\beta}$ with an additional factor $f_{\rm sky}$.

We split the theoretical, lensed $C_\ell$ into two contributions, as in Eq.\ (\ref{eq:split}),
\be
C^{ij}_\ell = C^{\rm lens}_\ell(z_i,z_j) =  C_\ell(z_i,z_j) + A_L \de C_\ell(z_i,z_j) \, , \label{eq:fisher1}
\ee
where $A_L$ is an (artificially introduced) amplitude of the lensing contribution, with a physical value of $A_L=1$.
We forecast the precision with which we can measure $A_L$. For this forecast we keep all cosmological parameters fixed, which is another optimistic assumption as parameter degeneracies can increase the actually achievable error bar on $A_L$. However, here we are only interested in an order of magnitude estimate. The derivative of the lensed $C_\ell$ with respect to $A_L$ is simply $\de C_\ell$. In this case, the Fisher matrix (\ref{eq:fishgen}) has only one element, which is given by
\be
F_{A_L A_L} = \sigma_{A_L}^{-2} = \sum_\ell \frac{2\ell+1}{2} \sum_{a,b,c,d} \left[ (\de C^{ab}_\ell) (C^{-1}_\ell)^{bc} (\de C^{cd}_\ell) (C^{-1}_\ell)^{da} \right] 
  \equiv \sum_\ell \frac{2\ell+1}{2} T_\ell \, . \label{eq:fisherAL}
\ee

We can obtain an estimate of the size of the effect by using the fact that the matrix $C_{\mu\nu} = C^{ij}_\ell$ is largest on the diagonal $i=j$, with much smaller off-diagonal terms (if the redshift bins are not too small). We can then write it as
\be
C_{\mu\nu} = \sqrt{C_{\mu\mu}}\sqrt{C_{\nu\nu}} \left(\delta_{\mu\nu} + \frac{\check{C}_{\mu\nu}}{\sqrt{C_{\mu\mu} C_{\nu\nu}}} \right) 
= \sqrt{C_{\mu\mu}}\sqrt{C_{\nu\nu}} \left(\delta_{\mu\nu} + \epsilon_{\mu\nu} \right) \, ,
\ee
where $\check{C}_{\mu\nu}$ contains only the off-diagonal terms, so that each element is much smaller than the diagonal elements in the denominator.
The two terms outside the bracket can be considered as diagonal matrices that are trivial to invert, while the matrix in brackets can be inverted with the help of a series expansion, $(1+\epsilon)^{-1} = 1 - \epsilon + O(\epsilon^2)$. The approximate inverse then becomes
\be
C_{\mu\nu}^{-1} \approx  \frac{ \delta_{\mu\nu} - \epsilon_{\mu\nu} }{\sqrt{C_{\mu\mu}}\sqrt{C_{\nu\nu}}}
= \frac{\delta_{\mu\nu}}{C_{\mu\mu}} - \frac{\check{C}_{\mu\nu}}{C_{\mu\mu}C_{\nu\nu}} \, .
\ee
We now use the fact that the non-lensed signal $C_\ell$ is to a good approximation diagonal, and that in addition we can keep the diagonal terms in $\de C_{\mu\nu}$ as they are highly subdominant, so that we can set $\check{C}_{\mu\nu} \approx \de C_{\mu\nu}$. The contributions $T_\ell$ to the Fisher matrix (\ref{eq:fisherAL}) then become
\begin{eqnarray} \label{eq.doublesum-in}
T_\ell &\equiv& \sum_{a,b,c,d} \left[ (\de C^{ab}_\ell) (C^{-1}_\ell)^{bc} (\de C^{cd}_\ell) (C^{-1}_\ell)^{da} \right] \\
&\approx&\sum_{a,b,c,d} \left[ 
(\de C^{ab}_\ell) \left( \frac{\delta_{bc} } {C_\ell^{bb} } - \frac{ \de C^{bc}_\ell }{ C_{bb} C_{cc} }  \right)
(\de C^{cd}_\ell) \left( \frac{\delta_{da} } {C_\ell^{dd} } - \frac{ \de C^{da}_\ell }{ C_{dd} C_{aa} }  \right)
\right] \\
&\approx& \sum_{a,c} \left[ 
\frac{\de C_\ell^{ac} \de C_{\ell}^{ca}}{C_\ell^{aa} C_\ell^{cc}} \label{eq:doublesum}
\right] + \ldots
\end{eqnarray}
where we only kept the leading order term and neglected terms of order $(\de C_\ell/C_\ell)^3$. This approximation shows that the effective error also for the off-diagonal terms ($a \neq c$), which are dominated by the lensing contribution, is given by the diagonal spectra, $C_\ell^{aa}$, which are much larger. Hence, the contribution from the off-diagonal terms to the Fisher matrix will be relatively small even though the relative contribution from the lensing is large for those elements.

 In Fig.\ \ref{fig:lensingsignal} we see that $\de C_\ell / C_\ell$ is of the order of $10^{-3}$ for $z_1=z_2$. If we just use this constant value on the diagonal and neglecting cross-correlations, we find as a rough approximation
\be
T_\ell \approx  n_z \left(\frac{\de C_\ell}{C_\ell} \right)^2 \, ,
\ee
and thus, using additionally that $\sum \ell \approx \lmax^2/2$,
\be
\sigma_{A_L} \sim \frac{C_\ell}{\de C_\ell} \frac{\sqrt{2}}{\lmax {\sqrt{n_z}}} \, ,
\ee
where $n_z$ denotes the number of (independent) redshift bins that we consider, and $\lmax$ the maximal value of $\ell$ that we use. For $\lmax = 2000$ and $n_z \approx 40$ (considering only the redshifts $z\in[2,6]$ as the signal is much smaller at low redshifts) we find $\sigma_{A_L} \approx 0.1$, in other words a detection should be possible -- and conversely, the second order lensing contribution to the IM power spectrum needs to be taken into account when analysing such a data set.

\begin{figure}[ht]
\begin{center}
\includegraphics[width=.9\textwidth]{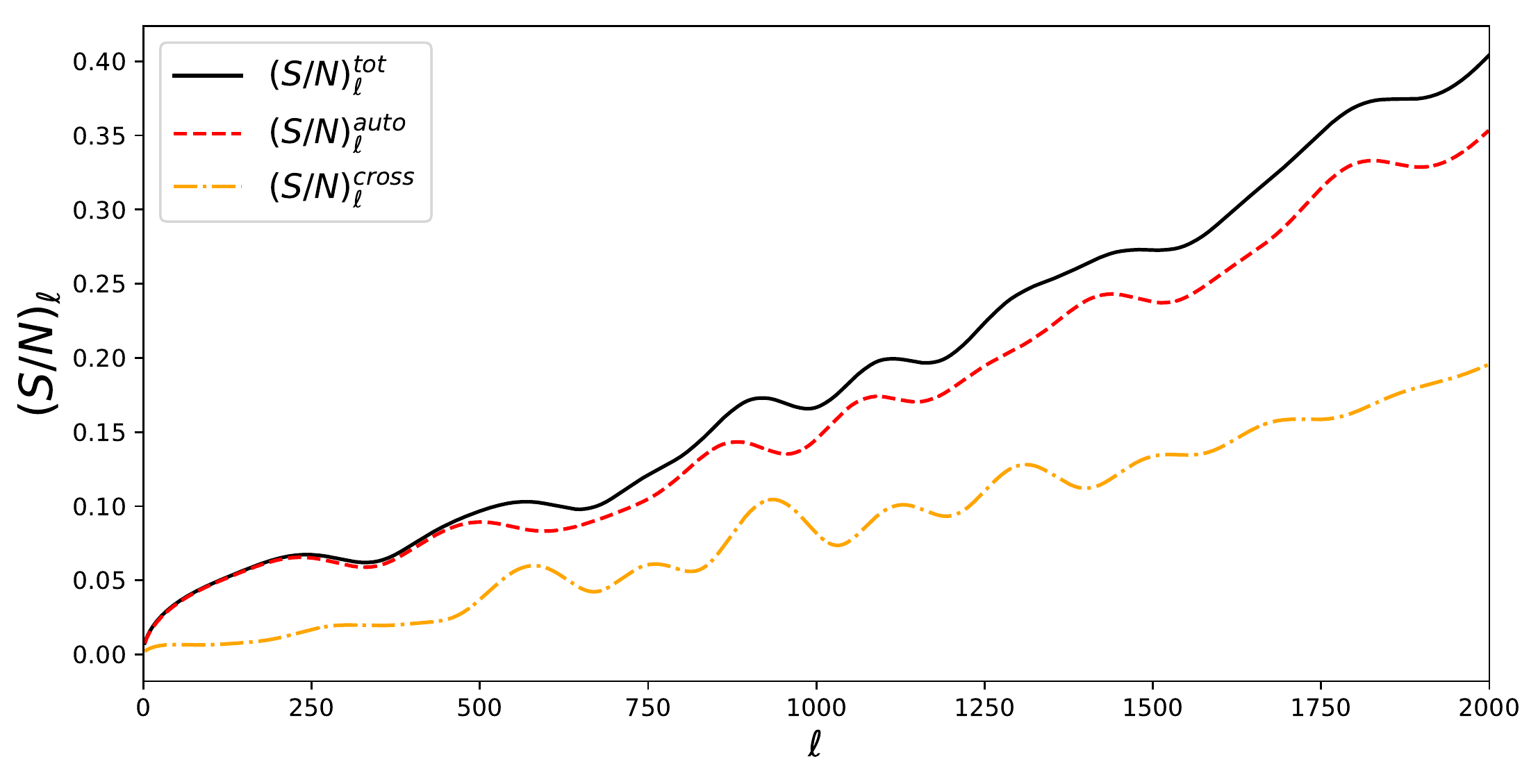}
\end{center}
\caption{Signal-to-noise ratio as a function of $\ell$, computed as the square root of the $\ell$-dependent terms in the Fisher matrix for $A$. The black solid line corresponds to the total signal-to-noise, while the red dashed line indicates the redshift auto-correlation terms and the orange dotted line indicates the redshift cross-correlation terms.}
\label{fig:SN}
\end{figure}
We note, a posteriori, that neglecting the higher order terms in (\ref{eq:doublesum}) is justified as long as $\sqrt{n_z} \ll C_\ell/\de C_\ell$, which is indeed the case here. We further note that since the double sum in Eq.\ (\ref{eq:doublesum}) is over positive definite quantities, no cancellation due to changing signs can occur.

To have a more precise estimate of the error on $A_L$, we compute the Fisher matrix by using Eq. (\ref{eq:fisherAL}) with the approximation derived in Eqs. (\ref{eq.doublesum-in})-(\ref{eq:doublesum}). We obtain $F_{A_L A_L} = 95.4$, resulting in an error for $A_L$ given by $\sigma_{A_L} = 1.02 \times 10^{-1}$, which is extremely close to the rough estimate given above: $\sigma_{A_L} \approx 0.1$.
In other words we obtain a signal-to-noise $S/N = \sqrt{F_{A_L A_L}}=1/\sigma_{A_L} \simeq  9.77$.  This means that even after accounting for more realistic survey specifications, it may be possible to detect lensing from IM of the 21 cm line.

 We verify that the approximation of Eqs. (\ref{eq.doublesum-in})-(\ref{eq:doublesum}) holds by computing the Fisher matrix directly from Eq. (\ref{eq:fisherAL}),  without any further approximation. The result is $F_{A_L A_L} = 94.2 $, with  $\sigma_{A_L} = {0.10}  $, 
which implies that the approximation of Eqs. (\ref{eq.doublesum-in})-(\ref{eq:doublesum}) is  accurate to about $ 1\%$. For this reason we will always use the aforementioned approximated formulas for the following calculations.

To understand how the different terms in the Fisher matrix contribute to the signal-to-noise, we 
plot the square root of the $\ell$-dependent terms appearing in Eq. (\ref{eq:fisherAL})  in Fig. \ref{fig:SN}. More precisely
\be
(S/N)_\ell\equiv \sqrt{(2\ell+1)T_\ell/2}=\sqrt{\sum_{ac}\left(S_\ell(z_a,z_c)\right)^2}\,.
\ee

Clearly, the main contribution comes from the auto-power spectrum terms, even though they are less in number (41 auto-power spectrum terms against 1640 cross-power spectrum terms). After summing them over $\ell$, the auto-power spectra give a Fisher matrix of $F_{A_L A_L}^{\rm auto} =  74.0$ while the sum of the cross-power spectra is $F_{A_L A_L}^{\rm cross} =  21.4$.
This can be understood by comparing the relative size of the different terms in the Fisher matrix shown in Fig.~\ref{fig:lensingsignal}. The amplitude of the auto-power spectra is about one order of magnitude larger than that of the cross-power spectra. The same can be seen in Fig.~\ref{fig:lensingvsz}, where lensing drops rapidly with increasing separation between the redshifts $z_1$ and $z$.

\begin{figure}[ht]
\begin{center}
\includegraphics[width=.9\textwidth]{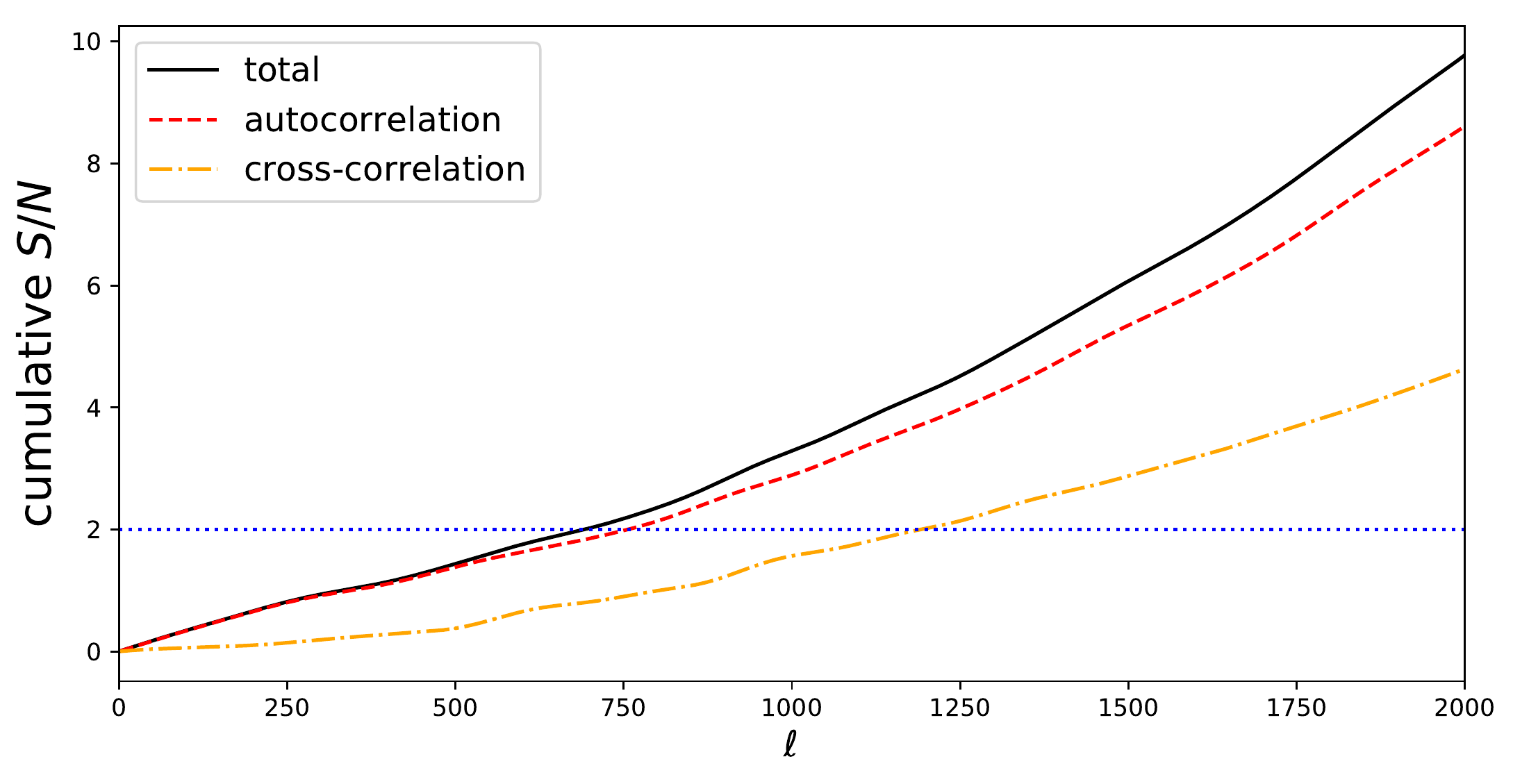}
\end{center}
\caption{Cumulative signal-to-noise as a function  of $\ell$. Black is the total signal, red-dashed is the contribution from redshift auto-power spectra alone while orange dot-dashed indicates the signal-to-noise from redshift cross-power spectra only.  Using  $\ell_{max} > 700$ results in a detection above 2$\sigma$.}
\label{fig:cumulativeS/N}
\end{figure}

To understand how much different $\ell$'s contribute to the total signal-to-noise, we also plot the cumulative  Fisher matrix as a function of $\ell$ in Fig. \ref{fig:cumulativeS/N}.
This shows clearly that large $\ell$ contribute most to the total signal-to-noise. As an example, the last $\sim 700$ $\ell$'s contribute about half of the total signal-to-noise.  It also shows again that auto-power spectra (red dashed lines in the plot) dominate the signal-to-noise if compared to the cross-power spectra terms (orange dot-dashed lines), which are negligible below $\ell \sim 300$, but add an extra $\Delta (S/N) \sim 1$ when using all $\ell$'s. 
We indicate with a blue dotted line the limit of a 2$\sigma$ detection: the threshold is crossed when using scales up to $\ell \gtrsim 700$. This also implies that neglecting lensing in IM will bias our results on cosmological parameters.

\begin{figure}[ht]
\begin{center}
\includegraphics[width=.9\textwidth]{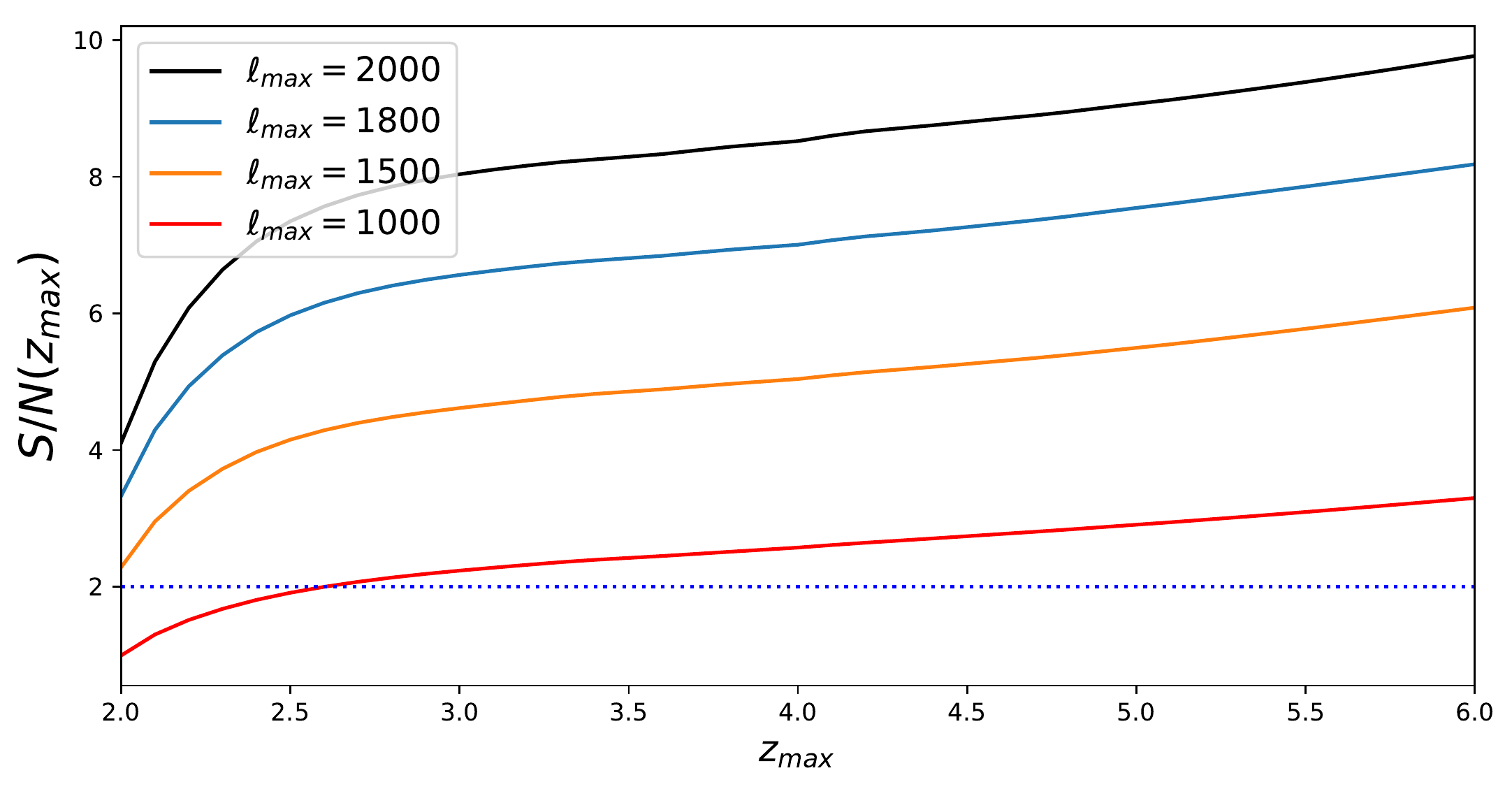}
\end{center}
\caption{Dependence of the signal-to-noise on the maximum redshift $z_{max}$ considered, i.e. the maximum redshift up to which we take the second sum of Eq.~\ref{eq:fisherAL}.  The black solid line is computed using all available $\ell$'s, i.e. with $\lmax = 2000$, while the blue line has $\lmax = 1800$, the orange line $\lmax = 1500$ and the red line $\lmax = 1000$.}
\label{fig:fisher-zdependence}
\end{figure}

In Fig.~\ref{fig:fisher-zdependence} we show how the total signal-to-noise depends on the maximum redshift bin considered.
 When $\ell$s up to $\ell_{max} = 2000$ are considered (black line), we obtain a $\sim$ 4$\sigma$ detection already when using only the first redshift bin at $z=2$. If instead we only use the first $1000$ $\ell$s, a 2$\sigma$ detection is reached at $z_{max} \sim 2.6$. 

 In Table~\ref{tab:Fisher} we summarize the contributions to the signal-to-noise of the IM lensing signal from the different lensing terms. Interestingly, the new post-Born term is not only of the same order as the terms 1+2, but it actually dominates the signal.

\begin{table}[htp]
\begin{center}
\begin{tabular}{|l|c|}
\hline
& S/N for $A_L$ \\
\hline
total, $(\lmax,z_{\max})=(2000,6)$ & 9.8\\
\hline
total with exact formula,  $(\ell_{\max},z_{\max}=(2000,6)$ & 9.7 \\
\hline
only 1st and 2nd terms, $(\ell_{\max},z_{\max})=(2000,6)$ & 5.1\\
\hline
only 3rd term, $(\ell_{\max},z_{\max})=(2000,6)$ & 8.3\\
\hline

auto-correlation terms only,  $(\ell_{\max},z_{\max}=(2000,6)$ & 8.6\\
\hline
cross-correlation terms only,  $(\ell_{\max},z_{\max}=(2000,6)$ & 4.6\\
\hline
up to $\ell_{\rm max} = 1000$,  $z_{\max}=6$ & 3.3\\
\hline
up to $z_{\rm max} = 4$,  $\ell_{\rm max} = 2000$ & 8.5\\
\hline
\end{tabular}
\end{center}
\label{tab:Fisher}
\caption{Signal-to-noise for different settings. Clearly, the highest $z$-bins and the high $\ell$'s are most relevant. Also, auto-correlations are more important than cross-correlations, and the third term contributes more than the sum of first and second one. The approximation differs from the exact computation by only 1\%.
}
\end{table}%

 \begin{figure}[ht]
\begin{center}
\includegraphics[width=.9\textwidth]{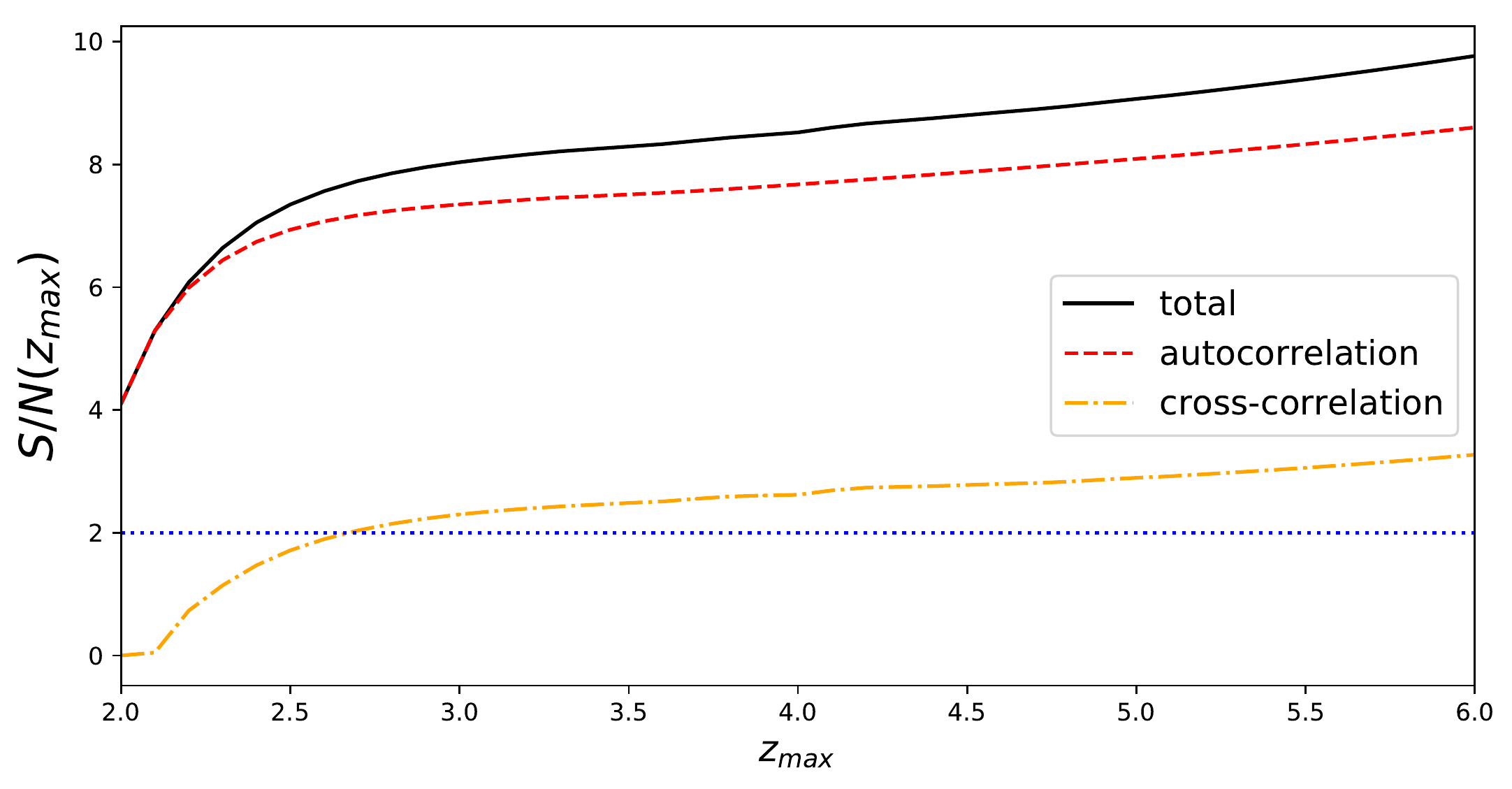}
\end{center}
\caption{Redshift dependence of the signal-to-noise, where the Fisher matrix element is computed by using the total signal (black solid line), or only auto-correlation terms (red dashed line), or only cross-correlation terms (orange dot-dashed line).}
\label{fig:fisher-autovscross-zdependence}
\end{figure}

\setcounter{equation}{0}

\section{Conclusions}
\label{Sec4}

In this paper we studied the lensing signal in intensity mapping (IM).  Like for the CMB, and contrary to number counts, there is no lensing contribution at first order, but at second and higher order lensing affects the power spectrum of both IM and the CMB.  At second order, however, there are more contributions in IM than in the CMB since neglecting the correlation between the lensing potential and the intensity fluctuations is no longer a good approximation. We have computed the corrections to the IM power spectrum by Taylor-expanding in the deflection angle up to second order. We found a new `post-Born' term which is not only relevant, but actually dominates the signal-to-noise for lensing of the IM. 

The lensing signal is much smaller than the analogous signal in the CMB. The main reason for this is the fact that the IM power spectrum is very smooth with virtually no structure so that the re-distribution of power due to lensing is a very small effect. Nevertheless, since we can consider IM power spectra of different redshifts, $C_\ell(z_1,z_2)$, we can use many more spectra than in the CMB, which allows in principle to overcome the smallness of the effect and to achieve a signal-to-noise of order 10. This shows that the lensing signal for such configurations has to be included and can affect the determination of cosmological parameters.

In this work we have considered intensity maps at redshift $z<z_{\rm reion}\sim 8$, i.e.\ after reionisation. Here we think of unresolved galaxies and proto-galaxies.  One could also study the signal before reionisation, $z \gtrsim 15$, where one would
mainly see density fluctuations. At redshifts in between, $8<z<15$, the signal is probably dominated by patchy reionisation and cannot be related in a simple way to structure formation but instead contains information about the reionisation process, which is interesting on its own right. IM at even higher redshifts can in principle also be used to study the density fluctuations of baryons in the `dark ages'.

In the future we also plan to study the effects of second and third order redshift space distortions (RSD) on intensity mapping spectra. The theoretical expressions are given in Eq.~(\ref{e:DHi3}) and we will analyse their amplitude and their spectral shape in future work. Since 21 cm intensity maps have excellent spectral resolution, higher order RSD should leave a characteristic imprint when we study very slim redshift bins. Increasing the number of redshift bins can in principle also help to increase the lensing signal-to-noise. For such very slim redshift bins, correlations between bins will come not only from the lensing signal but also from density and RSD and therefore be much larger.

It would also be interesting to study the effect of IM lensing on parameter estimation. As we have shown for number counts in the past, this can on the one side lift degeneracies and allow us to test General Relativity~\cite{Montanari:2015rga}, while on the other hand neglecting lensing can significantly bias parameter estimation~\cite{Cardona:2016qxn}.

\setcounter{equation}{0}

\section*{Acknowledgements}
It is a pleasure to thank Stefano Camera, Enea Di Dio, Giuseppe Fanizza, Basundhara Ghosh, Julien Lesgourgues, Kavilan Moodley, Hamsa Padmanabhan, Alkistis Pourtsidou and Nils Sch\"oneberg for helpful discussions.  
We acknowledge financial support from the Swiss National Science Foundation. This work was supported by a grant from the Swiss National Supercomputing Centre (CSCS) under project ID s710. The computations were performed on the Baobab cluster of the University of Geneva, and on Piz Daint at CSCS.

\vspace{2.5cm}

\appendix
\section{Derivation of the new lensing terms for HI intensity mapping\label{a:derCl-new}}
Starting from \eqref{e:DHlens}  we derive the result \eqref{e:Cellens} in some detail.
We do not repeat the derivation of \eqref{e:CMB2} which is the result obtained when neglecting correlations between the lensing potential and $\DH$, as this derivation can be found in books~\cite{RuthBook}. The new terms, which we obtain in addition, are those which correlate $\phi$ and $\DH$ or $\Psi$ and $\DH$.  In the correlation of the first and the third term of Eq.\ \eqref{e:DHlens}, $\l \DH(z_1)(\nabla^a\phi\nabla^b\phi\nabla_a\nabla_b\DH)(z_2)\re$, no such term survives since only vectors can be formed in this way, and these have vanishing expectation value due to statistical isotropy. However, the square of the second term contains in addition to the contribution given in \eqref{e:CMB2} a term coming from correlating $\phi$ with $\DH$. We compute it in the flat sky approximation. The 2d Fourier transform of $\nabla^a\phi\nabla_a\DH$ is
\be
{\rm FT}[\nabla^a\phi\nabla_a\DH](z_1,\bell) = \frac{-1}{2\pi}\int d^2\ell_1(\bell-\bell_1)_a\ell_1^a\phi(\bell-\bell_1)\DH(\bell_1) \,.
\ee
Correlating it with the corresponding expression at $z_2$ we obtain 
\bea
&& \langle {\rm FT}[\nabla^a\phi\nabla_a\DH](z_1,\bell) {\rm FT}[\nabla^a\phi\nabla_a\DH]^*(z_2,\bell')\rangle =  \nonumber\\  && \hspace*{3cm}
\de^2(\bell-\bell')\int\frac{d^2\ell_1}{(2\pi)^2}[\bell_1\cd(\bell-\bell_1)]^2C^\phi_{|\bell-\bell_1|}(z_1,z_2)C_{\ell_1}(z_1,z_2) +\nonumber \\
&&  \frac{1}{(2\pi)^2}\int d^2\ell_1d^2\ell_2(\bell-\bell_1)_a\ell_1^a(\bell'-\bell_2)_b\ell_2^b \l\phi(\bell-\bell_1,z_1)\DH(\bell_2,z_2)\re\l\phi(\bell'-\bell_2,z_2)\DH(\bell_1,z_1)\re \nonumber \\
&&  ~= ~ T_1 + T_2 \,.
\eea
The first term $T_1$ is the second term of \eqref{e:CMB2}, which also contributes to CMB lensing, but the term $T_2$ which, correlates  $\phi$ with $\DH$, is new.
Using $\l a(\bell)b^*(\bell')\re=\de^2(\bell-\bell')C^{ab}_\ell$ we find
\bea
T_2=\de^2(\bell-\bell')\int\frac{d^2\ell_1}{(2\pi)^2}[\bell_1\cd(\bell-\bell_1)]^2C^{\phi\DH}_{|\bell-\bell_1|}(z_1,z_2)C^{\DH\phi}_{\ell_1}(z_1,z_2) \,,
\eea
which contributes  the fourth term of \eqref{e:Cellens}. 

To obtain the third term we correlate the last term of \eqref{e:DHlens} with $\DH$.
The Fourier transform of this last term is
\bea
F_3(\bell_1,z_1) ~\equiv~{\rm FT}\left[-2\!\int_0^{r(z)}\!\!dr\frac{r(z)-r}{r(z)r} \left(\nabla^b\phi\nabla_b \nabla^a\Psi\right)(z)
\!\nabla_a\DH(z_1) \right](\bell_1)~=  \hspace{1cm} 
~~ && \nonumber\\
\frac{-2}{(2\pi)^2}\int_0^{r_1}\!\!dr\frac{r_1-r}{r_1r}\int\!\! d^2\ell d^2\ell'\ell^b\ell'_b\ell'_a(\bell_1-\bell-\bell')^a\phi(\bell,z)\Psi(\bell',z)\DH(\bell_1-\bell-\bell',z_1)  \,, && \label{a:F3}
\eea
where $r=r(z)$ and $r_1=r(z_1)$. Taking the expectation value of this expression multiplied by
$\DH(\bell_2,z_2)$ using Wick's theorem and statistical isotropy, only one term contributes,
\bea
\l F_3^*(\bell_1,z_1)\DH(\bell_2,z_2)\re &=& \de(\bell_1-\bell_2)\frac{-2}{(2\pi)^2}\int dr\frac{r_1-r}{r_1r}\int d^2\ell (\bell\cd\bell_2)^2C_{\ell_2}^{\Psi\DH}(z,z_2)C_\ell^{\phi\DH}(z,z_1) \nonumber\\
&=&  -2\de(\bell_1-\bell_2)\int dr\frac{r_1-r}{r_1r}\ell_2^2C_{\ell_2}^{\Psi\DH}(z,z_2)R^{\phi\DH}(z,z_1)   \,,
\label{ea:F3DH}
\eea 
where we have introduced
\be
R^{\phi\DH}(z,z_1) =\frac{1}{4\pi}\int_0^\infty d\ell \ell^3C_\ell^{\phi\DH}(z,z_1) \,.
\ee
We of course also have to take into account the symmetric term, 
\be
\l \DH(\bell_2,z_1)F^*_3(\bell_1,z_2)\re = -2 \delta(\bell_1-\bell_2) \int_0^{r_2} dr\frac{r_2-r}{r_2r}\ell_2^2C_{\ell_2}^{\Psi\DH}(z,z_1)R^{\phi\DH}(z,z_2)
\ee

We now re-express the Bardeen potential $\Psi$ using the Poisson equation. (This is only valid within $\La$CDM cosmology. If we have clustering dark energy or a modified theory of gravity, this simplifications are no longer valid.)
$$ 
\ell^2\Psi(\bell,z) = -r^2(z)\De_2\Psi(\bell,z) \simeq - r^2(z)\De\Psi(\bell,z) \,.
$$
Here $\De_2$ denotes the 2-dimensional spatial Laplacian in the plane normal to the radial direction and for the last $\simeq$ sign we use the fact that after integration over $r$ the radial derivatives become sub-leading boundary terms, which we can neglect.
Now for $\de_m$ the matter density contrast in comoving gauge we have
$$
\De\Psi =4\pi Ga^2\rho_m\de_m = \frac{3}{2}\Om_mH_0^2a^{-1}\de_m \,,
$$
so that
\be\label{ea:Clpside)}
\ell_2^2C_{\ell_2}^{\Psi\DH}(z,z_2) \simeq -\frac{3}{2} r^2(z)(1+z)\Om_mH_0^2C_{\ell_2}^{\de_m\DH}(z,z_2) \,.
\ee
Again, this equation becomes correct once integrated over $r$. Inserting this in \eqref{ea:F3DH} and adding the corresponding term with $z_1\leftrightarrow z_2$ we obtain the third term of \eqref{e:Cellens}.

\section{The computation of $\de C_{\ell}[3]$ }
\begin{figure}
\begin{center}
\includegraphics[scale=0.6]{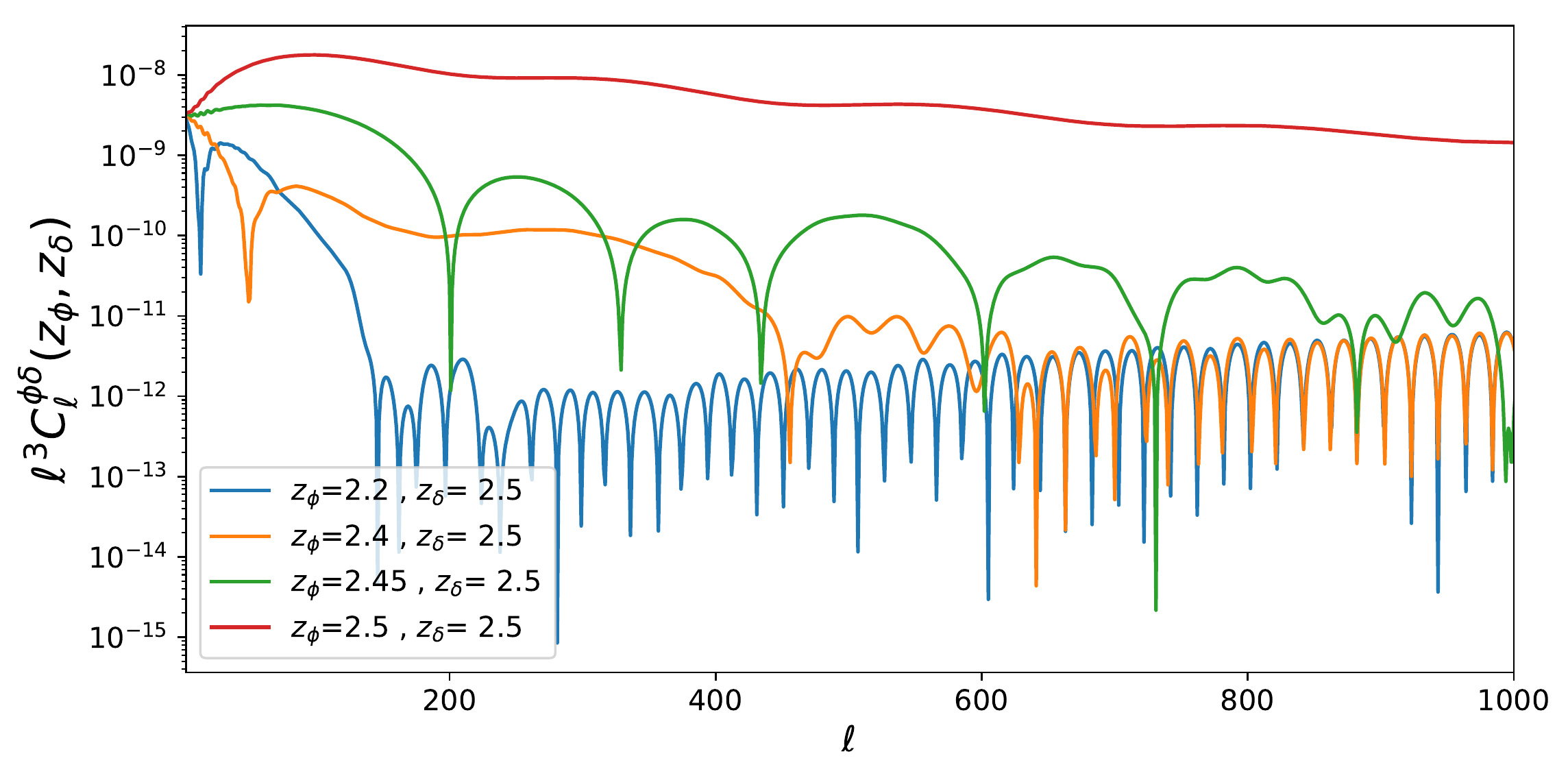}
\end{center}
\caption{$\ell^3 C_{\ell}^{\phi \delta}(z_\phi,z_\delta)$ for different redshift combinations with Dirac-$\delta$ window function.}
\label{fig:Clphidelta}
\end{figure}

In this appendix, we explain in detail the  computation of $\delta C_{\ell[3]}$. To calculate $R^{\phi \DH}(z_\phi,z_\De)$ that appears in $\delta C_{\ell[3]}$  given in Eq.~\ref{eq.lensing-terms}, we cannot use the Limber approximation since it vanishes in this approximation for $z_\phi<z_\De$. To compute $R^{\phi \Delta}$ we should integrate $C_{\ell}^{\phi \DH}$ over $\ell$ up to infinity, but in practice we can cut the integration once the integral has converged. In Table~\ref{table:rphidelta-lmax}, we give $\ell_{\max}$ up to which we integrate $C_{\ell}^{\phi \DH}$ for different redshift separations. We choose the range of integration as a function of the difference between $z_\phi$ and $z_\Delta$. The reason is that as the redshift difference increases, the integrand decays more quickly as a function of $\ell$, and spurious oscillations appear that we need to exclude from the integration. This is well visible in Fig.\ \ref{fig:Clphidelta}, which shows $\ell ^3 C_{\ell}^{\phi \DH}(z_\phi, z_\De) $ for different redshift separations.

\begin{table}[htp]
\begin{center}
\begin{tabular}{|c|c|}
\hline
redshift separation & $\ell_{\max}$\\
\hline
$z_\phi = z_\De$  & 2000  \\
\hline
$z_\phi=z_\De-0.05$ & 600 \\
\hline
$z_\phi=z_\De -0.1$ & 400 \\
\hline
$z_\phi<z_\De -0.1$ & 100 \\
\hline
\end{tabular}
\end{center}
\caption{\label{table:rphidelta-lmax}The upper limit of integration over $C_{\ell}^{\phi \DH}$ for computing $R^{\phi \delta}$ as a function of redshift separation.}
\end{table}%

\section{Halofit}
\begin{figure}[h]
\begin{center}
\includegraphics[scale=.6]{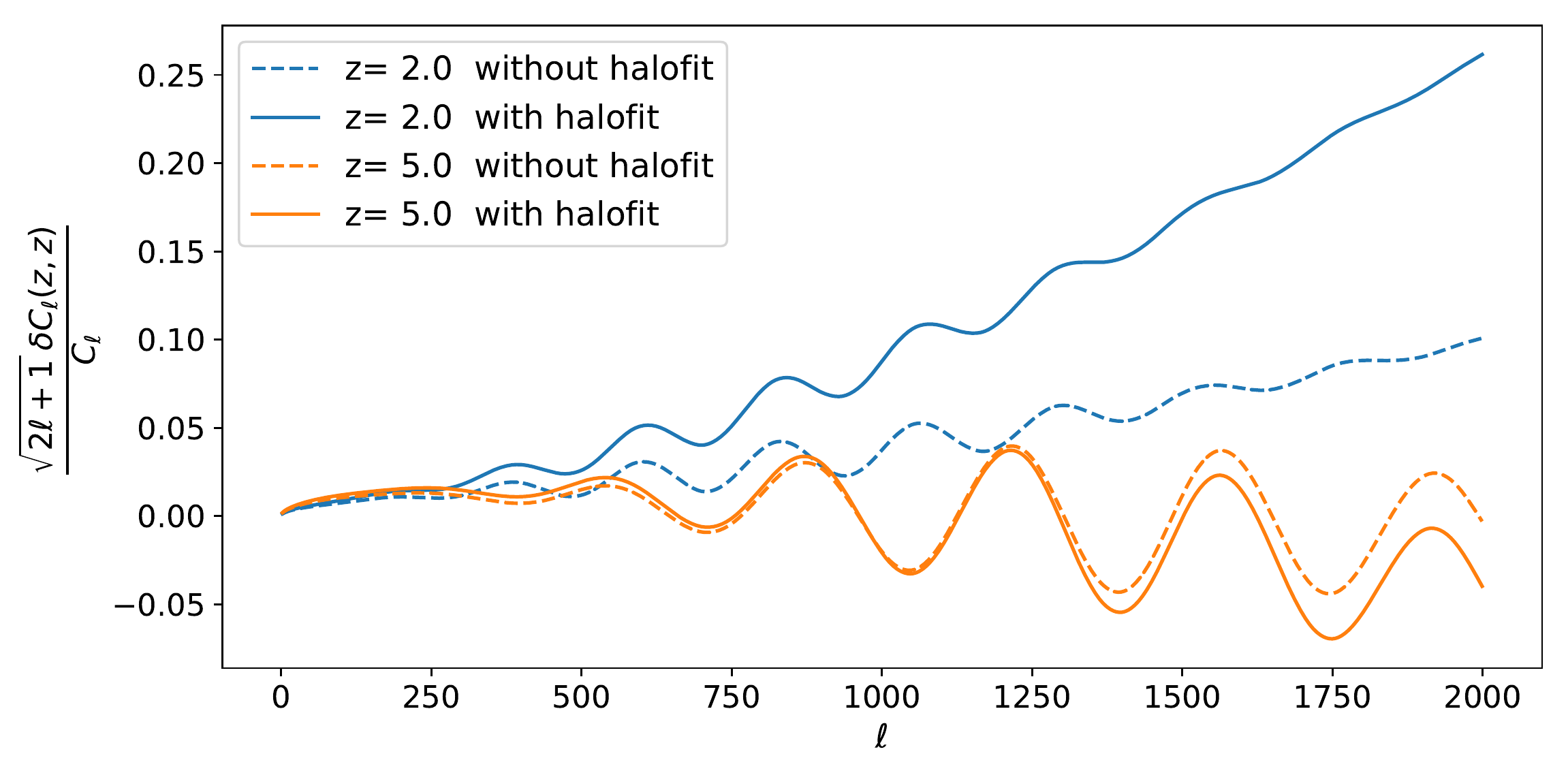}
\end{center}
\caption{The lensing signal, $\frac{\sqrt{2\ell +1} \, \delta C_{\ell}(z,z) }{C_{\ell}}$, computed with halofit (solid lines) and without halofit (dashed lines). The blue lines correspond to $z=2$ while the orange ones to $z=5$. When using halofit the lensing signal increases up to $200 \%$ for $z=2$ and up to $150\%$ for $z=5$.}
\label{fig:halofit}
\end{figure}

In our calculations we use halofit to compute power spectra.  This is a simple analytical transformation on the linear power spectrum which provides a good fit to N-body simulations~\cite{Takahashi:2012em}.  In Fig.~\ref{fig:halofit} we show the effect of halofit on lensing signal, $\frac{\sqrt{2\ell +1} \, \delta C_{\ell}(z,z) }{C_{\ell}}$, for $z=2$ and $z=5$. Taking into account halofit increases our lensing signal significantly. In particular at redshift $z=2$, the signal increases by up to a factor $2$, and  at redshift $z=5$, it increases by up to a factor $1.5$. At higher redshift halofit is less important since a large portion of the lensing integral is in the linear regime.

In $C_\ell^{\de_m\De}(z,z')$ we need the halofit power spectrum at different redshifts which we simply approximate as
\be
P_\de(k,z,z')\simeq \sqrt{P_\de(k,z)P_\de(k,z')} \,.
\ee
This is the so called perfectly coherent approximation which we has not been tested with numerical simulations. However, since the diagonal terms $z\simeq z'$ largely dominate, we are confident that this will not lead to a large error in our final results.


\section{Cancellation of $\de {C_\ell}_{[1]}$ and $ \de {C_\ell}_{[2]}$ in IM and in the CMB}\label{app:cancellation}
\begin{figure}[h]
\begin{center}
\includegraphics[scale=0.6]{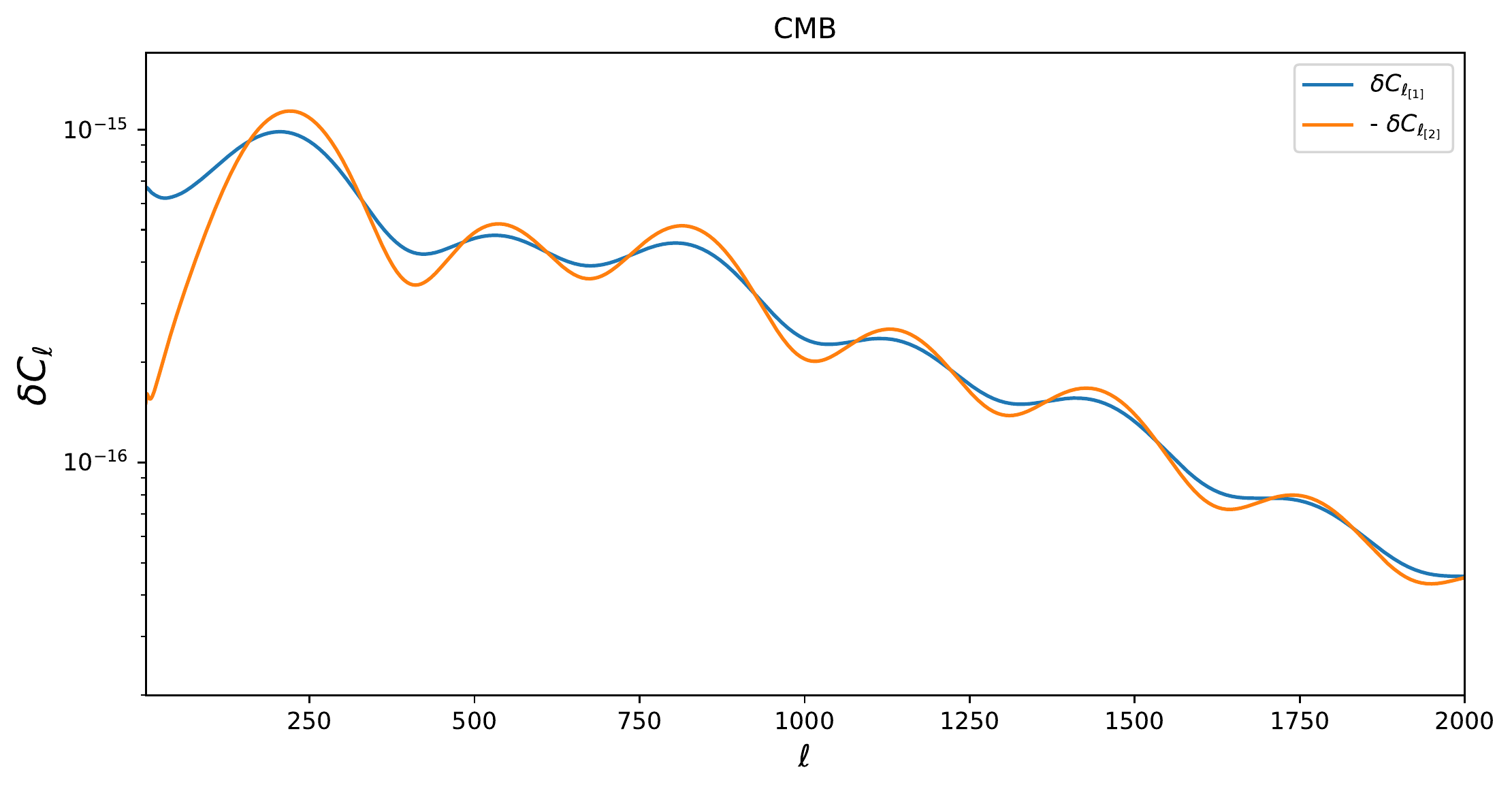}
\includegraphics[scale=0.6]{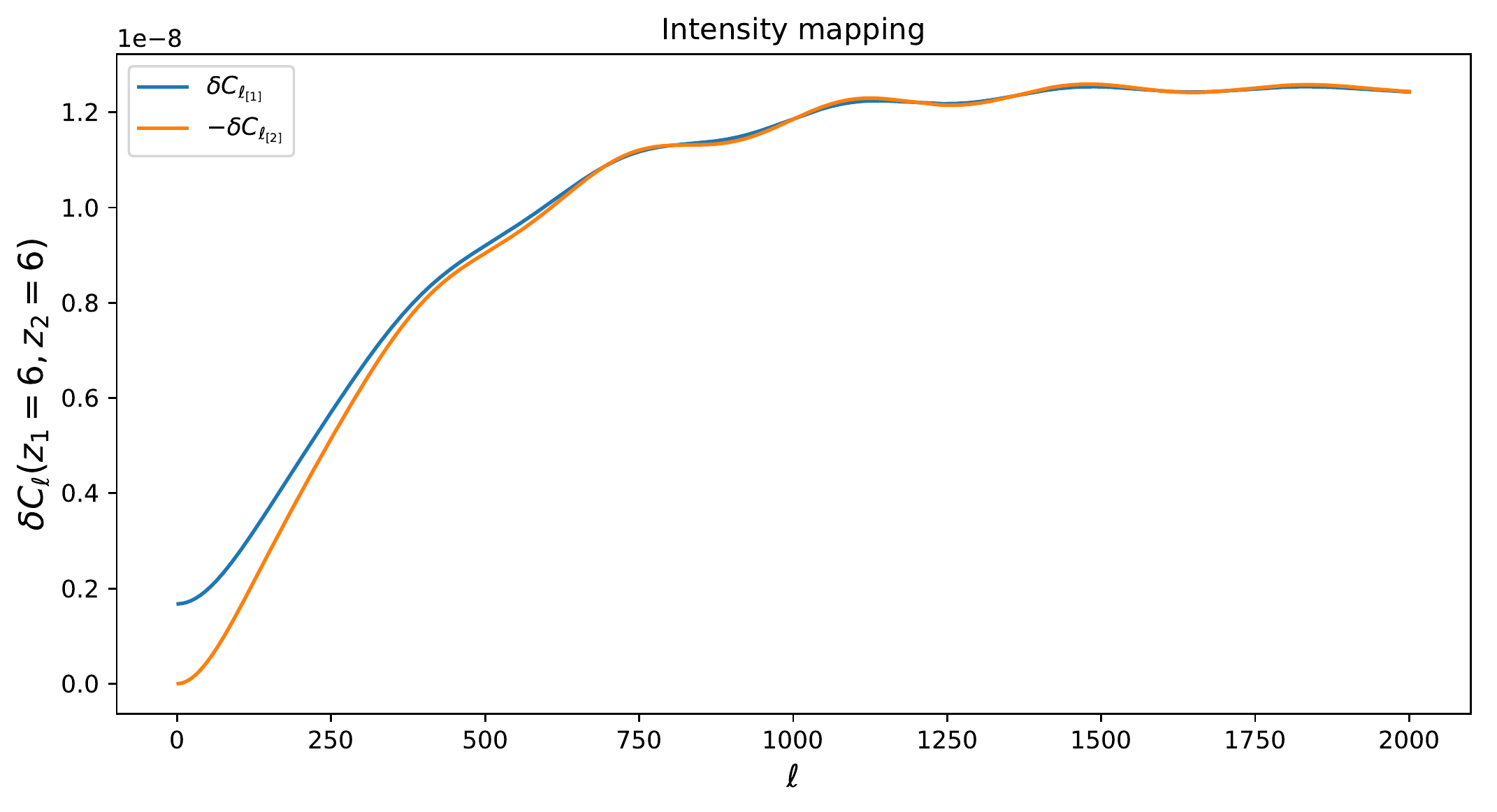}
\end{center}
\caption{Second order lensing terms $\delta C_{\ell_{[1]}}$ and $-\delta C_{\ell_{[2]}}$ for the CMB (top plot), and IM (bottom plot). The cancellation of $\delta C_{\ell_{[1]}}$ with $\delta C_{\ell_{[2]}}$ happens both for the CMB and intensity mapping, but for the CMB this cancellation is less precise due to the prominent oscillations of the CMB power spectrum. }
\label{fig:cancellation}
\end{figure}

\begin{figure}[h]
\begin{center}
\includegraphics[scale=0.6]{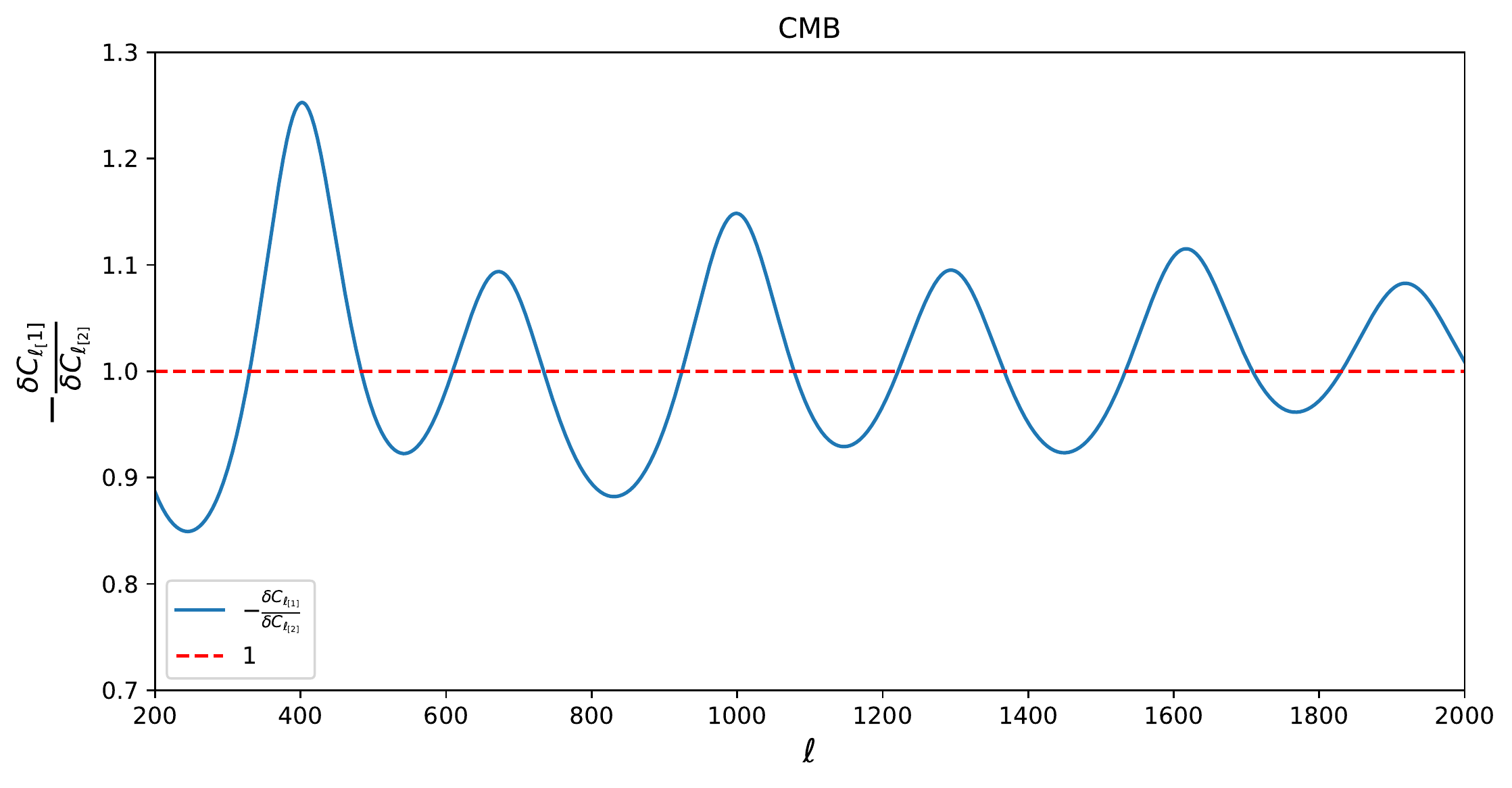}
\includegraphics[scale=0.6]{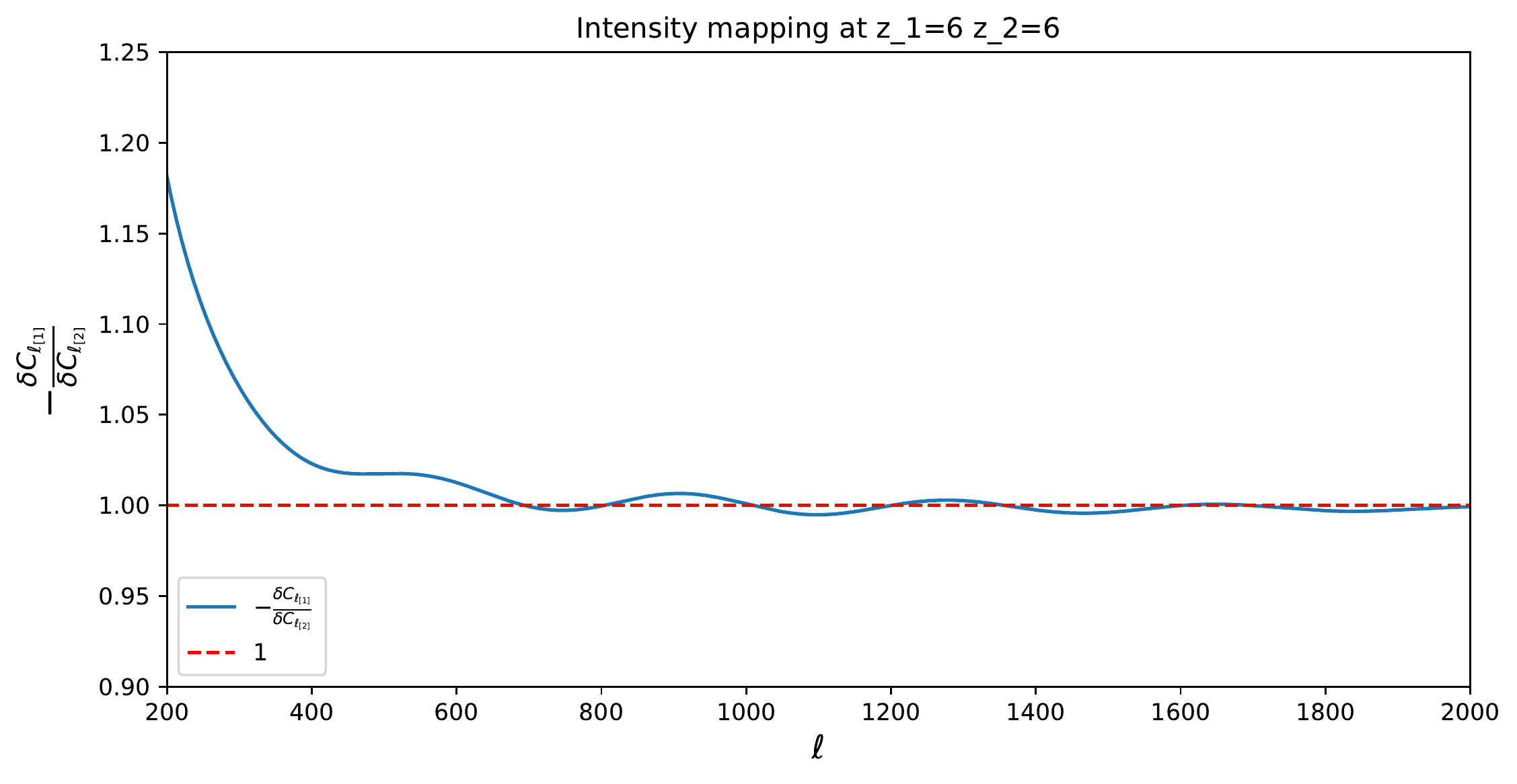}
\end{center}
\caption{The upper plot shows the term $-{\delta C_{\ell_{[1]}}}/{\delta C_{\ell_{[2]}}}$ for CMB, and the bottom plot shows the same for intensity mapping. The ratio $-{\delta C_{\ell_{[1]}}}/{\delta C_{\ell_{[2]}}}$ is almost 1 for intensity mapping, especially for high $\ell$ but for CMB the ratio $-{\delta C_{\ell_{[1]}}}/{\delta C_{\ell_{[2]}}}$ has large oscillations, due to the oscillations of the CMB power spectrum. This renders the CMB lensing signal so significant. }
\label{fig:cancellationdevide}
\end{figure}

As we know from Section~\ref{sec:lensedCl}, the terms $\de {C_\ell}_{[1]}$ and $ \de {C_\ell}_{[2]}$ of Eq.~(\ref{eq.lensing-terms}) are present also in the CMB. In the case of IM lensing, they nearly cancel each other, as we show in Figs.~\ref{fig:lensingterms} and~\ref{fig:lensingterms2}. In this appendix, we show that this cancellation also happens to some extent for the CMB, and we explain why the CMB lensing signal is nevertheless so large. In Fig.~\ref{fig:cancellation}, we show the behavior of $\de {C_\ell}_{[1]}$ (blue lines) and $-\de {C_\ell}_{[2]}$ (orange lines) for the CMB (top plot) and for intensity mapping (bottom plot).   The CMB lensing terms have larger oscillations, and therefore $\de {C_\ell}_{[1]}$ and $\de {C_\ell}_{[2]}$ cancel much less precisely.  To see this, we plot the ratio $-\de {C_\ell}_{[1]} / \de {C_\ell}_{[2]}$ in Fig.~\ref{fig:cancellationdevide}; a ratio of 1 implies total cancellation. For the CMB, the ratio, $-\de {C_\ell}_{[1]} / \de {C_\ell}_{[2]}$ , has large oscillations around 1, while for IM it is almost equal to 1 except for $\ell<600$. 
This is due to the smoothness of the intensity mapping power spectrum compared to the CMB power spectrum, see  Fig.~\ref{fig:powerspectrum}. If the $C_{\ell}$'s are almost constant we can take them out of the integral in the definition of $\de {C_\ell}_{[1]}$ in Eq.~\eqref{eq.lensing-terms}, which leads to 
 $$\de {C_\ell}_{[1]} \simeq C_{\ell} \int \frac{d^2\ell'}{(2\pi)^2}[\bell'\cd(\bell-\bell')]^2  C^\phi_{\ell'}(z_1,z_2)\simeq \ell^2 C_{\ell} R^{\phi}= -\de {C_\ell}_{[2]} \,.$$


\section{The windowed $\de {C_\ell}_{[3]}$} \label{app:third-term-derivation}
As explained in Section~\ref{s:num},  we observe a windowed power spectrum,
\begin{align}
&\langle \DH^{*{\rm lens}}(\bell_1,z_1) \DH^{\rm lens}(\bell_2,z_2) \rangle^{\rm obs} =\nonumber \\ &\int dz'W(z',z_2) \int dz'' W(z'',z_1) \langle{\Delta}_{\mathrm{HI}}^{*{\rm lens}}(\bell_1,z''){\Delta}_{\mathrm{HI}}^{\rm lens}(\bell_2,z') \rangle
\end{align}
where $W(z',z)$ is a normalized window function around $z$, and ${\Delta}_{\mathrm{HI}}(\bell,z)$ on the right hand side represents a not-windowed  fluctuation at redshift z. Using \eqref{eq.lensing-terms} we find that the windowed term $\de {C_\ell}_{[3]}$ is given by
\begin{align}
&\langle  \de {C_\ell}_{[3]} \rangle^{\rm obs}
= 3\Omega_{m0} H_0^2 \int_{0}^{\infty} dz'' W(z'',z_1) \int_0^{z''}dz\frac{(r''-r)r}{r''}\frac{(1+z)}{H(z)}
\nonumber \\
&\times
\int_{0}^{\infty} dz'W(z',z_2)C_{\ell_2}^{\delta_m \De }(z,z')
\,
R^{\phi\Delta}(z,z'')\,. \label{eq:deltaCl3-general}
\end{align}

Let us choose a normalized top-hat window functions, where the width of the function is denoted by $\DW$.
In this case Eq.~(\ref{eq:deltaCl3-general}) becomes
\bea  \label{Cl3-exact}
\langle  \de {C_\ell}_{[3]} \rangle^{\rm obs}&= \frac{ 3 \Omega_{m0} H_0^2}{(\DW)^2} \int_{z_1-\DW/2}^{z_1+\DW/2} dz'' \int_{0}^{z''} dz\,  B(z,z'') R^{\phi\Delta}(z,z'') \nonumber\\
& \times \int_{z_2-\DW/2}^{z_2+\DW/2} dz' C_{\ell}^{\Delta \delta }(z',z)\,,
\eea
where $B(z,z'') = (1+z)(r''-r)r/[H(z)r'']$.

The first term $\propto R^{\phi\Delta}(z,z'')$ is already an integral over $z''$ and therefore very smooth. The additional window integral will probably not have a significant effect on it. We therefore can to a good approximation neglect the first integral and fix $z''$ to be equal to $z_1$.
\bea
\langle  \de {C_\ell}_{[3]} \rangle^{\rm obs} \simeq \frac{ 3 \Omega_{m0} H_0^2}{\DW}  \int_{0}^{z_1} dz\,  B(z,z_1) R^{\phi\Delta}(z,z_1) \int_{z_2-\DW/2}^{z_2+\DW/2} dz' C_{l}^{\Delta \delta }(z',z)\,.
\label{appendix:upper-limit-approximation}
\eea

\begin{figure}[h]
\begin{center}
\includegraphics[scale=0.6]{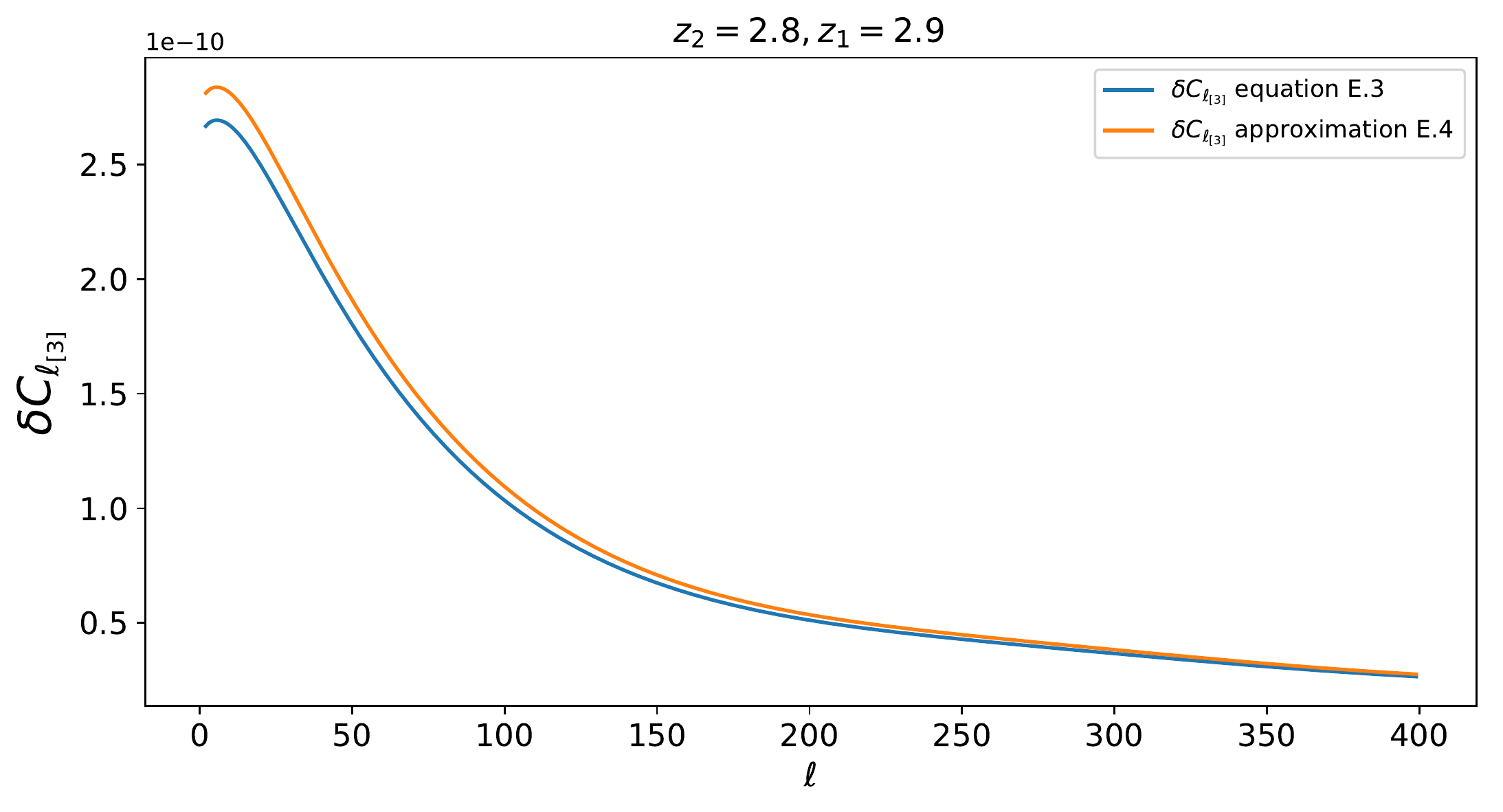}
\caption{\label{fig:thirdterm-approx-z5}  $\delta C_{\ell_{[3]}}(z_2=2.8, z_1 = 2.9)$ computed with no approximation from equation \eqref{Cl3-exact} (blue line) and computed with the approximation \eqref{appendix:upper-limit-approximation} (orange line). }
\end{center}
\end{figure}

In Fig.~\ref{fig:thirdterm-approx-z5} we show how good these approximations are.

\section{ {\sc class} settings} \label{app:CLASS-settings}
In this appendix we briefly explain the settings used to compute power spectra with the {\sc class} code~\cite{Blas:2011rf,DiDio:2013bqa}. To compute $\delta C_{\ell [1]}$ and $\delta C_{\ell [2]}$, we use a top-hat window function with full width $\Delta z =0.1$ with \texttt{l\_max\_lss} = 5000. We set the Limber parameters to \texttt{l\_switch\_limber\_for\_nc\_local\_over\_z} = 10000 and 
\texttt{l\_switch\_limber\_for\_nc\_los\_over\_z} = 2000. To compute $\delta C_{\ell [3]}$, we need  Dirac-$\delta$ window functions. We noticed that for this window function (and also for other choices) 
 the power spectra $C_{\ell}^{\delta \delta}(z_1,z_2)$ for a fixed redshift pair $(z_1,z_2)$ depend on the mean redshift of the redshift list (\texttt{selection\_mean} in  {\sc class}). To avoid this dependence we need to set the parameter \texttt{k\_max\_tau0\_over\_l\_max} = 15. We also use \texttt{l\_linstep} = 10 and \texttt{l\_logstep} = 1.07 to sample more frequently in $\ell$ space with respect to the default {\sc class} values.

\bibliographystyle{JHEP}
\bibliography{biblio}
\end{document}